\documentclass[preprint,prd,aps,nofootinbib,10pt]{revtex4-2}
\usepackage{amsmath}
\usepackage{graphicx}
\usepackage{subcaption}
\usepackage{tikz}
\usepackage{diagbox}
\usepackage{slashed}
\usepackage{float}
\usepackage{color}
\usepackage{makecell}
\usepackage{hyperref}

\begin{document}
\title{Strong decays of $\psi(3770)$, $\psi(4040)$ and $\psi(4160)$ in four relativistic wave-function representations}

\author{Wen-Yuan Ke$^{1,2,3}$,
Wan-Li Ju$^{1,2,3,4}$,
Tianhong Wang$^5$,
Qiang Li$^6$,
Guo-Li Wang$^{1,2,3}$\footnote{corresponding author}}
\affiliation{$^1$ Department of Physics, Hebei University, Baoding 071002, China\\
$^2$ Hebei Key Laboratory of High-Precision Computation and Application of Quantum Field Theory, Baoding 071002, China\\
$^3$ Hebei Research Center of the Basic Discipline for Computational Physics, Baoding 071002, China\\
$^4$ Department of Physics, University of Alberta, Edmonton AB T6G 2J1, Canada\\
$^5$ School of Physics, Harbin Institute of Technology, Harbin 150001, China\\
$^6$ School of Physical Science and Technology, Northwestern Polytechnical University, Xi'an 710072, China
}

\begin{abstract}
Strong decays provide a direct test of the relativistic structure of heavy
quarkonium wave functions. We study the OZI-allowed hadronic decays of the
vector charmonia $\psi(3770)$, $\psi(4040)$ and $\psi(4160)$ within the
Salpeter equation, using four relativistic wave-function representations
$\varphi_1$--$\varphi_4$ that differ in the $S$-, $P$- and $D$-wave
components they retain independently, and the revised ${}^3P_0$ model. The
quark-pair creation strength is fixed once, by reproducing the measured
$\psi(3770)\to D\bar D$ width, and is not readjusted afterwards, so that the
results for the two higher states are parameter-free predictions. The
representation $\varphi_2$ predicts $94^{+34}_{-23}$~MeV for $\psi(4040)$
and $100^{+23}_{-20}$~MeV for $\psi(4160)$, in agreement with the measured
$84.5\pm12.3$ and $69\pm10$~MeV, and it is the only one whose predicted
width ratios are compatible with the data. By contrast, $\varphi_1$
overshoots $\psi(4040)$ by $2.5\sigma$, $\varphi_3$ gives the largest widths
and is disfavoured at the $5\sigma$ level, and $\varphi_4$, which has no
independent $S$-wave component, underestimates $\psi(3770)$ by about
$2\sigma$ and cannot be applied to final states containing vector mesons.
These results show that a successful representation must contain
independent $S$, $P$ and $D$ components, the $P$-wave terms providing the
dominant relativistic correction, while the nodal structure of the wave
functions accounts for the large uncertainties of the predictions. The
mixing angles extracted from $\varphi_2$, $7.92^{\circ}$, $25.8^{\circ}$ and
$26.3^{\circ}$, agree with the values obtained in coupled-channel models.
\end{abstract}

\maketitle

\section{Introduction}
\label{sec:intro}

The properties of vector charmonia above the open-charm threshold have attracted continuing experimental and theoretical interest since their discovery. $\psi(3770)$ was first observed in 1977 \cite{Rapidis:1977}; it is predominantly a $1^3D_1$ state with a small $2^3S_1$ admixture \cite{Rosner:2001nm}, and its main decay modes are $D^0\bar{D}^0$ and $D^+D^-$ \cite{PDG}. The experimentally measured total width of $\psi(3770)$ is $27.2\pm1.0$~MeV, with the $D\bar D$ channel accounting for $93\%$; the corresponding sum of partial widths is $25.3^{+2.0}_{-2.4}$~MeV \cite{PDG}, which provides a key constraint for theoretical models. $\psi(4040)$ and $\psi(4160)$ were discovered in $e^+e^-$ annihilation experiments and have more complex decay patterns involving several open-charm channels \cite{PDG,BES:2007zwq,BES:2006psi}.

All these states are $S-D$ mixed states, and the mixing angles reported by different authors often differ considerably. The traditional approach is to describe the particles in terms of nonrelativistic $S$-wave and $D$-wave eigenstates and to combine them by artificially introducing a mixing angle, which is subsequently determined by fitting experimental data. Such a procedure has limited predictive power: the mixing is an external input rather than a result of the dynamics, so that different fits and different processes often lead to different mixing angles. 
The mixing angle of $\psi(3770)$ quoted in different works ranges from
about $5^{\circ}$ all the way to $40^{\circ}$
\cite{Barnes:2005pb,Chao:2008,Badalian:2009}. The identification of the
$\psi(4040)$ and $\psi(4160)$ states has itself remained a subject of
debate \cite{Close:1995}; traditionally they are assigned as
$3\,{}^3S_1$ and $2\,{}^3D_1$ \cite{Godfrey:1985,Barnes:2005pb}, and the
$3S$--$2D$ mixing between them may significantly affect their decay
properties \cite{Chao:2008,Badalian:2009}. In contrast to the
artificial-mixing approach, the coupled-channel analyses attribute the
mixing to meson-loop effects. The $3S$--$2D$ mixing angles quoted for
$\psi(4040)$ and $\psi(4160)$ nevertheless differ considerably: for
instance, $\theta_1=7^{\circ}$ and $\theta_2=10^{\circ}$ are found in
Ref.~\cite{Man:2025}, $23.4^{\circ}$ in Ref.~\cite{Man:2024}, and
$28^{\circ}$ and $29^{\circ}$ in Ref.~\cite{Badalian}.

The decay properties of $\psi(4040)$ present a particularly interesting puzzle. Despite the ample phase space, the $D\bar D$ channel is strongly suppressed while $D\bar D^*$ and $D^*\bar D^*$ dominates, and $D_s\bar D_s$ remains appreciable. This hierarchy clearly deviates from the naive phase-space expectation \cite{Piotrowska:2018,Ding:2024zy}. If $\psi(4040)$ is interpreted as a $3\,{}^3S_1$ charmonium state, the node of its radial wave function offers a natural explanation \cite{Barnes:2005pb}: the $D\bar D$ overlap integral receives positive and negative contributions of comparable size from different momentum regions, leading to strong destructive interference, whereas the $D\bar D^*$, $D^*\bar D^*$ and $D_s\bar D_s$ channels do not experience the same suppression.

It should be noted that the $2S\text{-}1D$ and $3S\text{-}2D$ mixings mentioned above are inherently 
non-relativistic descriptions. Within this framework, hadrons are characterized by the non-relativistic 
spectroscopic notation $^{2S+1}L_J$, where $S$- and $D$-waves correspond to orbital angular momentum eigenvalues 
of $0$ and $2$, respectively. Alternatively, we have developed a relativistic formalism based on $J^{P(C)}$ 
quantum numbers \cite{Wang:2005}. In this framework, a general wave-function representation for a hadron is constructed from 
its $J^P$ quantum numbers. Consequently, the resulting relativistic wave functions are no longer conventional 
pure states; for instance, a $1^-$ meson generally contains $S$-, $D$-, and even $P$-wave components. Furthermore, 
such mixing is not introduced in an ad hoc manner, but is instead dictated by the underlying relativistic 
dynamical equation, the Salpeter equation \cite{Salpeter:1952}, that the wave function satisfies. Meanwhile, in our relativistic 
formalism, the non-relativistic parity relation $P = (-1)^{L+S}$ and charge conjugation parity relation 
$C = (-1)^{L+1}$ cease to be valid \cite{Wang:2022cxy}.

The resulting general wave function $\varphi_1$ for a $1^-$ meson provides an excellent description of the 
properties of $S$-wave-dominated states, such as the mass spectra of heavy quarkonia \cite{Wang:2005,Chang:2010},  radiative transitions \cite{Pei:2023}, and dilepton annihilation decays of $1^{--}$ quarkonium \cite{Ke:2026gnk}, etc. However, when applied to the dilepton decay of $D$-wave-dominated quarkonium, it fails to reproduce the 
experimental results \cite{Ke:2026gnk}. We noted that, besides $\varphi_1$, three additional wave-function representations are 
possible. In a recent work~ \cite{Ke:2026gnk}, we therefore performed a comparative study of all four admissible 
wave functions for $1^{--}$ quarkonia. We solved the respective Salpeter equations and investigated the mass 
spectra and dilepton decay widths of charmonium and bottomonium states. The mass spectra alone cannot distinguish the four representations, while the dilepton widths provide a clear criterion: only the $\varphi_2$ representation simultaneously reproduces the PDG values of the $S$-wave dominated states ($J/\psi$, $\psi(3686)$, $\psi(4040)$) and of the $D$-wave dominated states ($\psi(3770)$, $\psi(4160)$). The dilepton process is most sensitive to the mixing structure of the wave function and therefore gives the most stringent discrimination. Specifically, $\varphi_2$ gives $\Gamma(\psi(3770)\to e^+e^-)=0.231^{+0.091}_{-0.067}$~keV, in agreement with the experimental value $0.262\pm0.018$~keV \cite{PDG}. The $\varphi_1$ representation underestimates the dilepton width of the $D$-wave states, $\varphi_3$ systematically overestimates the $S$-wave widths, and $\varphi_4$ cannot produce $S$-wave dominated states at all.

A further analysis of the wave function $\varphi_2$ reveals that the $1^{--}$ states are not simply $S$-$D$ mixed state but rather an $S$-$P$-$D$ mixture. For charmonium states, the resulting $S:P:D$ ratios are:
$1 : 0.0639 : 0.281$ for the $J/\psi$, 
$1 : 0.180 : 0.117$ for the $\psi(3686)$, 
$0.139 : 0.221 : 1$ for the $\psi(3770)$, 
$1 : 0.252 : 0.483$ for the $\psi(4040)$, 
and $0.495 : 0.339 : 1$ for the $\psi(4160)$ \cite{Ke:2026gnk}.  The $P$-wave component carries a non-negligible relativistic correction. If it is discarded and the traditional effective mixing angle is defined only from the $S$- and $D$-wave ratio, we obtained $\theta_{\psi(3770)}=7.92^{\circ}$, $\theta_{\psi(4040)}=25.8^\circ$ and $\theta_{\psi(4160)}=26.3^{\circ}$ \cite{Ke:2026gnk}. 
The former is in agreement with the small mixing angle reported in Refs.~\cite{Ding:1991vu,smallangle1,smallangle2}, 
whereas the latter is consistent with the large mixing angle obtained in Refs.~\cite{Chao:2008,Badalian:2009}.

In the present paper we extend this comparison from electromagnetic processes to strong decays and test whether the same choice of wave function still holds. We apply the modified ${}^3P_0$ model to calculate the OZI-allowed strong decays of $1^{--}$ charmonia. Since the experimental data for $\psi(3770)$ are the most complete, we fix the quark pair creation strength $\gamma_u=\gamma_d=0.527$ from the partial width of $\psi(3770)\to D\bar D$, and take $\gamma_s=0.35$ from the dynamical mass suppression relation $\gamma_s/\gamma_u=m_u/m_s$. The predictions for $\psi(4040)$ and $\psi(4160)$ therefore require no additional free parameters. The strong decay analysis independently points to the $\varphi_2$ representation, in agreement with the conclusion of the dilepton analysis; the electromagnetic and strong decay observables give the same choice of the wave function, showing that $\varphi_2$ captures the dominant relativistic dynamics of vector quarkonia.

The paper is organized as follows, Sec.~\ref{sec:theory} introduces the wave function picture, the four wave function representations, and the revised  ${}^3P_0$ model; Sec.~\ref{sec:results} presents the numerical results, analyzes the node mechanism, and compares with other theories and experiment; Sec.~\ref{sec:conclusion} gives the summary and conclusions; Appendix~\ref{app:salpeter} introduces the Salpeter equation;  Appendix~\ref{app:coeff} gives the positive-energy coefficients, and the normalization conditions.

\section{Theoretical framework} \label{sec:theory}

\subsection{Wave function representations of the $1^{-}$ meson}

In the non-relativistic limit, a vector meson is purely an $S$-wave, i.e., a $^3S_1$ state, with wave function 
\begin{align}
\left(M + \slashed{P}\right)\slashed{\epsilon}f(-{q}_{_\perp}^2),
\end{align} 
where, $\epsilon$ is the polarization vector of the meson, $P$ and $M$ denote the meson momentum and mass, respectively, while the radial wave function 
$f$ depends on the squared relative momentum between the quark and antiquark within the meson, 
$-{q}_{_\perp}^2 = \vec{q}^2$. The relative momentum $q$ is related to the quark momenta 
by: $p_1 = \alpha_1 P + q$ for the quark and $p_2 = \alpha_2 P - q$ for the antiquark, where 
$\alpha_1 = m_1/(m_1+m_2)$ and $\alpha_2 = m_2/(m_1+m_2)$, with $m_1$ and $m_2$ being the constituent 
masses of the quark and antiquark, respectively. The relative momentum $q$ is decomposed into components 
parallel and perpendicular to $P$: $q_{\parallel} = (P \!\cdot\! q / M^2) P$ and 
$q_{_\perp} = q - q_{\parallel}$. In the rest frame of the meson, under the instantaneous approximation, 
one has $q_{_\perp} = (0, \vec{q})$.

In our earlier work~\cite{Wang:2005}, we constructed a general wave-function representation 
\begin{align}\label{wavefull}
\varphi_{\text{1}}=\varphi_{1^{-}}^{\text{full}} = \epsilon \cdot q_{_\perp} \left( f_1 +  \frac{\slashed{P}}{M}f_2+ \frac{\slashed{q}_{\perp}}{M} f_3 + \frac{\slashed{P}\slashed{q}_{\perp}}{M^2} f_4 \right) + \left(M f_5 + \slashed{P} f_6\right)\slashed{\epsilon}+\left(  f_7 + {\frac{\slashed{P}}{M} f_8}\right){\slashed{\epsilon} \slashed{q}_\perp}
\end{align} that satisfies the $1^-$ quantum numbers within the instantaneous approximation  \cite{Ke:2026gnk}, where the radial wave functions $f_i$ ($i=1,2,...,8$) are functions of $-{q}_{_\perp}^2$. A comparison 
shows that terms $f_5$ and $f_6$ correspond to the non-relativistic wave function and represent $S$-wave 
contributions. Our analysis further reveals that terms $f_1$, $f_2$, $f_7$, and $f_8$ each contain a 
single factor of ${q}_{_\perp}$, identifying them as $P$-wave terms. Terms $f_3$ and $f_4$, on the other hand, contain 
two factors of ${q}_{_\perp}$ and therefore comprise both $S$-wave and $D$-wave components, specifically with the $S$-wave 
part given by $\frac{1}{3}q_{_\bot}^{2}\slashed{\epsilon}$ and the $D$-wave part by $(\epsilon\cdot q_{_\perp})\slashed{q}_{\perp}-\frac{1}{3}q_{_\bot}^{2}\slashed{\epsilon}$. It is thus evident that our constructed general wave 
function for a $1^-$ meson is an $S$-$P$-$D$ admixture. The relative proportions of different partial waves 
and the numerical solutions for the radial wave functions $f_1, f_2, \dots, f_8$ are obtained by solving 
the Salpeter equation with $\varphi_1$ as input. Appendix~A provides a brief overview of the Salpeter equation and the interaction potential employed. 

We note, however, that while $\varphi_1$ provides a 
general representation for $1^-$ mesons, it is not unique. As shown in Ref.~\cite{Ke:2026gnk}, three alternative 
forms of the relativistic wave function exist, corresponding to the following three cases: 
\begin{align} \label{wave}
& \varphi_{\text{2}}= \epsilon \cdot {q_{}}_{\perp} \left( f_1 + \frac{\slashed{P}}{M} f_2 + \frac{\slashed{q}_{\perp}}{M} f_3 + \frac{\slashed{P}\slashed{q}_{\perp}}{M^2} f_4 \right) + \left(M f_5 + \slashed{P} f_6\right)\slashed{\epsilon}, \nonumber \\
& \varphi_{\text{3}}= \epsilon \cdot {q_{}}_{\perp} \left(\frac{\slashed{q}_{\perp}}{M} f_3 + \frac{\slashed{P}\slashed{q}_{\perp}}{M^2} f_4 \right) +  \left(M f_5 + \slashed{P} f_6\right)\slashed{\epsilon}+\left(  f_7 + {\frac{\slashed{P}}{M} f_8}\right){\slashed{\epsilon} \slashed{q}_\perp}, \nonumber \\
& \varphi_{\text{4}}= \epsilon \cdot {q_{}}_{\perp} \left( f_1 + \frac{\slashed{P}}{M} f_2 + \frac{\slashed{q}_{\perp}}{M} f_3 + \frac{\slashed{P}\slashed{q}_{\perp}}{M^2} f_4 \right) +\left(f_7 + {\frac{\slashed{P}}{M} f_8}\right){\slashed{\epsilon} \slashed{q}_\perp}.
\end{align}
In fact, alternative forms of the $1^-$ meson wave function exist beyond the four representations 
discussed above. However, as noted in Ref.~\cite{Ke:2026gnk}, such alternative forms are either 
non-relativistic or can be subsumed into one of the above four categories, and are thus omitted 
from the present discussion.

Among the above wave-function representations, $\varphi_1$ contains eight radial wave functions, 
whereas each of the remaining three contains six, however, these radial wave functions are not all independent. 
The Salpeter equation comprises a set of four coupled equations, see Eq.~(\ref{Lorentz-covariant}) 
in the Appendix A, the last two of which, $\varphi^{+-} = \varphi^{-+} = 0$, constrain the number 
of independent radial wave functions. In each case, only four of them are independent, 
and we choose $f_3$, $f_4$, $f_5$, and $f_6$ as the independent ones. The other radial wave functions 
are determined in terms of these; specifically, for $\varphi_1$, one finds 
$$
f_1 = \frac{(w_1+w_2)f_3 q_{\perp}^2+2w_2 f_5 M^2}{M(m_1 w_2+m_2 w_1)},~
f_2 = \frac{(w_1-w_2)f_4 q_{\perp}^2-2w_2 f_6 M^2}{M(m_1 w_2+m_2 w_1)},~
f_7 = -\frac{f_5 M (w_1-w_2)}{m_1 w_2+m_2 w_1},~
f_8 = \frac{f_6 M (w_1+w_2)}{m_1 w_2+m_2 w_1}.
$$
Those corresponding relations for $\varphi_2$, $\varphi_3$ and $\varphi_4$, as well as the normalization 
conditions for these wave functions, are collected in Appendix~B.

\subsection{Solution of the Full Salpeter Equation}

Expressing the four $1^-$ meson wave functions $\varphi_i$ in terms of the four unknown independent radial wave 
functions and inserting them into the first two equations of the Salpeter equation set, Eq.~(\ref{Lorentz-covariant}), we evaluate 
the traces either directly or after multiplying both sides of the equation by an appropriate operator, 
such as $\slashed{P}$, $\slashed{q}_{\bot}$, $\slashed{\epsilon}$, etc. For each $\varphi_i$, the number of independent Dirac-matrix-free equations 
thus obtained is not two but four (see, e.g., Ref.~\cite{Wang:2019tqf}).
Consequently, we obtain four 
independent equations for the four unknown independent radial wave functions, which renders the full 
Salpeter equation solvable and yields the mass eigenvalues and numerical solutions for the 
radial wave functions of a $1^-$ meson. 
This procedure requires the positive-energy wave functions of the mesons, which are also needed 
later in the calculation of the transition amplitude. Their definition, Eq.~(\ref{posi}) in 
Appendix~A, shows that the positive-energy wave function is obtained by projecting the 
full meson wave function onto its positive-energy component. For the four $1^-$ wave functions 
$\varphi_i$, the corresponding positive-energy wave functions admit the unified representation
\begin{align}\label{positron}
\varphi^{++}_{_P}(q_{_\bot}) = &\, \epsilon\!\cdot\!q_{_\bot} \left(A_{1}+\frac{\slashed{P}}{M}A_{2}+\frac{\slashed{q}_{\bot}}{M}A_{3}+\frac{\slashed{P}\slashed{q}_{\bot}}{M^{2}}A_{4}\right)+ M\slashed{\epsilon}\left(A_{5}+\frac{\slashed{P}}{M}A_{6}+\frac{\slashed{q}_{\bot}}{M}A_{7}+\frac{\slashed{P}\slashed{q}_{\bot}}{M^{2}}A_{8}\right),
\end{align}
with the coefficients $A_i$ ($i=1,2,\dots,8$) varying 
across the four cases. Their specific expressions are shown in the Appendix B.

The eigenvalues and radial wave functions obtained from the solutions reveal that the first three sets 
of wave functions, i.e., $\varphi_1$, $\varphi_2$, and $\varphi_3$, yield qualitatively similar 
behaviors. While all solutions are $S$-$P$-$D$ admixtures, the dominant partial waves differ among 
them, with the solutions corresponding successively to states whose dominant components are $1S$, $2S$, 
$1D$, $3S$, $2D$, $\dots$. In contrast, for $\varphi_4$, although the solutions are again $S$-$P$-$D$ 
admixtures, they are uniformly D-wave-dominated, with its spectrum consisting successively of $1D$, 
$2D$, $3D$, $\dots$ states.

\begin{figure}[htbp]
\centering
\includegraphics[width=0.7\textwidth]{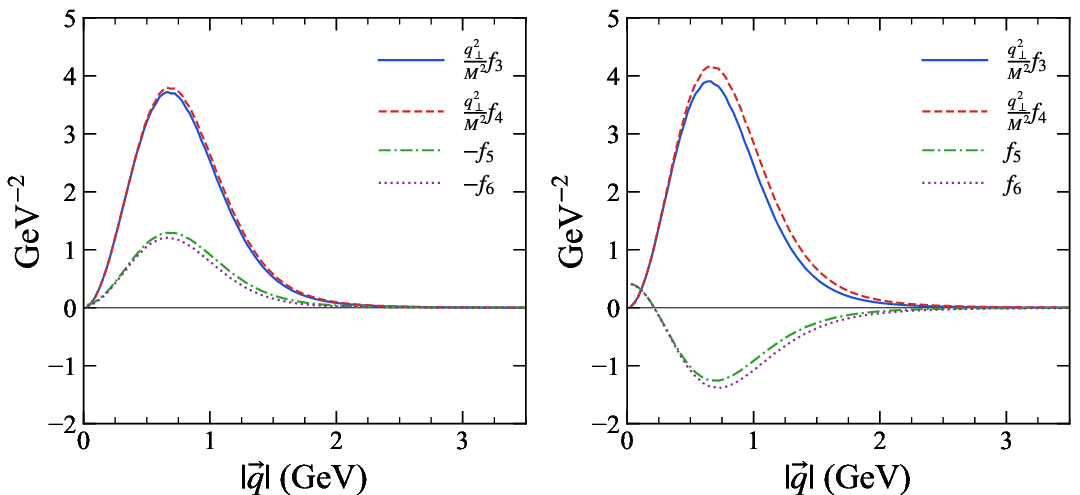}
\vspace{4mm}
\includegraphics[width=0.7\textwidth]{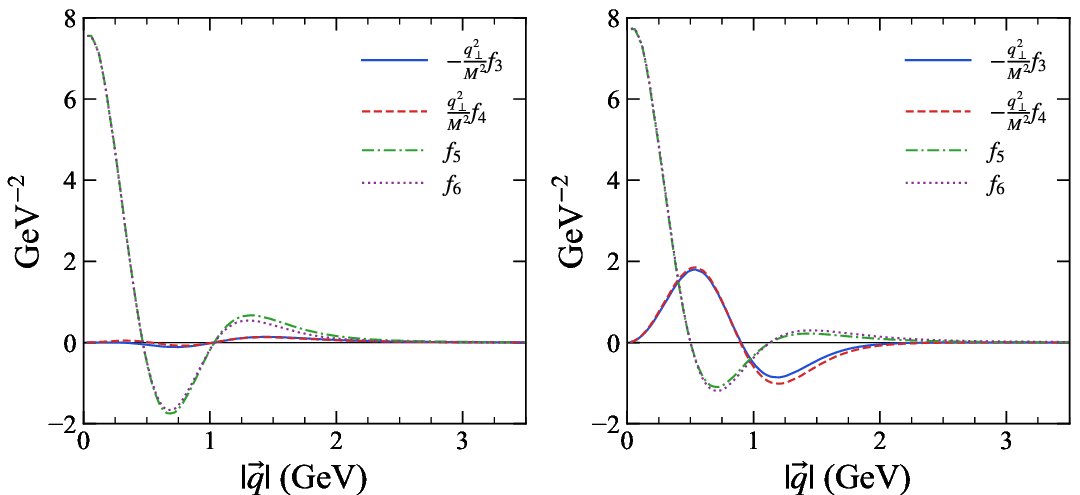}
\caption{The third and fourth solutions of $\varphi_1$ and $\varphi_2$. 
The top two panels show the third solutions, with dominant contributions from $1D$ wave functions; 
the bottom two panels show the fourth solutions, with dominant contributions from $3S$ wave functions. 
Thus, the upper panels display the radial wave functions of $\psi(3770)$ given by $\varphi_1$ 
and $\varphi_2$, while the lower panels display those of $\psi(4040)$.}\label{fig:wf}
\end{figure}

To illustrate the wave-function behavior, we examine the third and fourth solutions of $\varphi_1$ 
and $\varphi_2$ as representative examples. In Fig.~\ref{fig:wf}, the upper two panels display the third 
solutions of $\varphi_1$ and $\varphi_2$, respectively. It is evident that the radial wave functions 
$f_3$ and $f_4$ yield the dominant contributions. As noted previously, these components are predominantly 
$D$-wave; furthermore, the absence of nodes in the wave patterns indicates that they are $1D$-dominated 
states. Their predicted masses are $3773.6~\text{MeV}$ and $3773.3~\text{MeV}$ \cite{Ke:2026gnk}, respectively. 
Combining the wave-function structure with the mass values, we identify these states with the 
experimentally established $\psi(3770)$. The lower two panels present the fourth solutions of 
$\varphi_1$ and $\varphi_2$. Here, the radial wave functions $f_5$ and $f_6$ provide the largest 
contributions and are of $S$-wave character. Furthermore, the wave functions exhibit two nodes. Given their theoretical masses of 
$4051.4~\text{MeV}$ and $4052.2~\text{MeV}$, the fourth solutions of $\varphi_1$ and 
$\varphi_2$ are identified as the 3S-dominated state $\psi(4040)$. 

\subsection{The revised ${}^3P_0$ model for strong decays}

The OZI-allowed strong decays of a heavy quarkonium into two mesons can be described by the ${}^3P_0$ model \cite{Micu:1969,LeYaouanc:1973}, also called the quark pair creation model. In this model a light quark-antiquark pair with quantum numbers $J^{PC}=0^{++}$ (${}^3P_0$) is created from the vacuum. The interaction Hamiltonian is
\begin{equation}
H_I = 2m_q \gamma_q \int d^3x\, \bar\psi(x)\psi(x),
\end{equation}
where $m_q$ is the constituent mass of the created light quark $q$ and $\gamma_q$ is the dimensionless quark pair creation strength. It should be emphasized that the ${}^3P_0$ model was originally nonrelativistic: both the vertex and the transition amplitudes were constructed within the nonrelativistic quark model. We have extended the model to the relativistic case \cite{Fu:2012zzb}, that is, we use the relativistic wave functions obtained from the Salpeter equation with the same vertex structure, instead of nonrelativistic wave functions; the calculations of this paper are carried out within this extended framework, see the next subsection.

The total width of the $\psi(3770)$, together with the branching fraction and partial width of 
the $\psi(3770) \to D\bar{D}$ channel, has been well established by multiple experiments 
 \cite{PDG,BES:2006psi}. These measurements thus provide the input needed to fix the $\gamma_q$ parameter of the 
$^{3}P_{0}$ model. As shown in Table IV of Ref.~\cite{Ke:2026gnk}, among the wave-function 
representations, $\varphi_2$ gives the best description of $D$-wave-dominated states such as the 
$\psi(3770)$. We therefore adopt $\varphi_2$ in the present work and apply the relativistic 
$^{3}P_{0}$ model to compute the $\psi(3770) \to D\bar{D}$ decay. Fitting the calculated width 
to the data yields the quark-pair creation strength as
\begin{equation}\label{gamma-value}
\gamma_u = \gamma_d = {0.527}.
\end{equation}
This value is consistent with the typical range used in the ${}^3P_0$ model. It is then used directly for $\psi(4040)$ and $\psi(4160)$ without any further adjustment, so that the widths of these two states are genuine predictions. 

The treatment of the strange quark pair creation parameter $\gamma_s$ deserves particular attention, since it directly affects the predicted widths of the channels involving $D_s$ mesons. In contrast to the $\psi(3770)$ case, precise measurements of the $\psi(4040)$ and $\psi(4160)$ 
decays into $D^+_s D^-_s$ are absent because of limited event statistics. Consequently, 
$\gamma_s$ cannot be extracted from data. In the literature, $\gamma_s$ has been estimated in 
different ways. Assuming flavor universality of $\gamma_q$, one finds $\gamma_s = \gamma_u$; 
by contrast, if flavor universality is imposed on $g = 2m_q\gamma_q$ rather than on $\gamma_q$, 
one arrives at $\gamma_s = \gamma_u\, m_u/m_s$. Furthermore, because the $s$ quark is 
appreciably heavier than the $u$ and $d$ quarks, strange-quark-pair creation in the vacuum is 
kinematically disfavored, naturally suppressing $\gamma_s$ with respect to $\gamma_u$ and 
$\gamma_d$. Balancing the above theoretical inputs and physical considerations, in the present work we adopt the 
pragmatic choice 
\begin{equation}
\gamma_s = 0.35.
\end{equation}

\subsection{Decay amplitude and positive wave function of $0^-$ meson}
\label{sec:pipeline}

In the revised relativistic ${}^3P_0$ model, the transition amplitude of an OZI-allowed two-body strong decay is the overlap integral of the Salpeter wave functions of the three participating mesons with the ${}^3P_0$ quark pair creation vertex \cite{Fu:2012zzb}:
\begin{equation}\label{genamp}
\mathcal{M} = 2m_q\gamma_q \int \frac{d^3\vec q}{(2\pi)^3}\,
\mathrm{Tr}\Big[ \frac{\slashed P}{M} \varphi_{_P}^{++}(q_{_\perp})\,\frac{\slashed P}{M} \,
\bar\varphi_{_{P_{f_1}}}^{++}(q_{_{f_1\perp}})\,\bar\varphi_{_{P_{f_2}}}^{++}(q_{_{f_2\perp}}) \Big],
\end{equation}
where, $P$ and $M$ denote the momentum and mass of the initial decaying meson; 
$P_{f_1}$ and $q_{_{f_1\perp}}$ are, respectively, the total and internal relative momenta of the 
final meson containing the $\bar{c}$ quark, while $P_{f_2}$ and $q_{_{f_2\perp}}$ are the 
corresponding quantities for the final meson containing the $c$ quark. 
$\varphi^{++}$ stands for the positive-energy wave function, with the conjugate 
wave function defined as $\bar{\varphi} = \gamma^0 \varphi^\dagger \gamma^0$. 
In the spectator picture, the internal momenta of the initial and final mesons are related by
\begin{equation}
q_{_{f_1\perp}} = q_{_\perp} + \frac{m_c}{m_c+m_q} P_{f_1\perp},\qquad
q_{_{f_2\perp}} = q_{_\perp} - \frac{m_c}{m_c+m_q} P_{f_2\perp},
\end{equation}
where $m_q$ is the mass of the light quark (or light antiquark) belonging to the light 
quark–antiquark pair created out of the vacuum, and $m_c$ is 
the charm (anticharm) quark mass.

In the present work the initial state is a $1^-$ vector charmonium, and the final states consist 
of $1^-$ and $0^-$ mesons. The positive-energy wave function of the $1^-$ meson entering 
Eq.~(\ref{genamp}) was given above. We now quote the positive-energy wave function of the $0^-$ 
meson \cite{Ke:2024semi},
\begin{equation}
\begin{aligned}\label{positive 0-}
\varphi_{_P}^{++}(q_{\perp})=
\left( B_1 + B_2\frac{\slashed{P}}{M} + B_3\frac{\slashed{q}_{\perp}}{M} + B_4\frac{\slashed{P}\slashed{q}_{\perp}}{M^2} \right)\gamma^{5},
\end{aligned}
\end{equation}
where the $B_i$ ($i=1,2,3,4$) are scalar coefficients built out of the radial wave functions; 
their explicit form is collected in Appendix~B.

Inserting the positive-energy wave functions of the three mesons (initial and final) into 
Eq.~(\ref{genamp}) yields the transition amplitude by a straightforward calculation, the details of 
which we skip. Given the transition amplitude, the decay width reads
\begin{equation}
\Gamma = \frac{|\vec P_f|}{8\pi M^2} \, \overline{|\mathcal{M}|^2}=\frac{|\vec P_f|}{8\pi M^2} \frac{1}{2J+1} \sum |\mathcal{M}|^2,
\end{equation}
where $\vec P_f\equiv \vec P_{f_1}=-\vec P_{f_2}$ is the magnitude of the three-momentum of the final-state mesons, $J=1$ denotes the total angular momentum of the initial $1^-$ meson, 
and the sum runs over the polarizations of the initial and final states.

\section{Numerical results}
\label{sec:results}

In this section we first list the input parameters used in the calculation, then present the OZI-allowed strong decay widths of $\psi(3770)$, $\psi(4040)$, and $\psi(4160)$, analyze the node mechanism responsible for the suppression of the $\psi(4040)\to D\bar D$ channel, and estimate the theoretical uncertainties.

\subsection{Input parameters}

The types of decays considered in this paper are:
\begin{enumerate}
\item $1^- \to 0^- + 0^-$: $\psi \to D\bar D$, $\psi \to D_s\bar D_s$,
\item $1^- \to 1^- + 0^-$: $\psi \to D^*\bar D + c.c.$, $\psi \to D_s^* D_s + c.c.$,
\item $1^- \to 1^- + 1^-$: $\psi \to D^*\bar D^*$.
\end{enumerate}
The mesons considered here belong to two classes: $1^-$ vector mesons and $0^-$ pseudoscalar 
mesons. Pseudoscalar wave functions are comparatively simple: they all share a single functional 
form, and the Salpeter equations for different pseudoscalar mesons differ solely in the values of 
their parameters. Vector mesons, by contrast, are more involved. For $D$-wave-dominated charmonia, 
the wave function can be chosen from four possibilities, $\varphi_1$, $\varphi_2$, $\varphi_3$, 
and $\varphi_4$, each governed by a different Salpeter equation that must therefore be solved 
individually. For $S$-wave-dominated charmonia and heavy–light charmed mesons, three wave-function 
options, $\varphi_1$, $\varphi_2$, and $\varphi_3$, are available. Our input parameters are thus 
essentially those required by the Salpeter equation, namely the quark masses and the parameters of 
the modified Cornell potential. The explicit form of the modified Cornell potential and its parameters are detailed 
in Appendix~A; here we only list the values of input parameters in Table~\ref{tab:params}.

\begin{table}[ht]
\begin{center}
\caption{Input parameters.}\label{tab:params}
\setlength{\tabcolsep}{4pt}{
\begin{tabular}{l c c c c c c c c}
\hline
$m_c$ & $m_u$ & $m_d$ & $m_s$ & $\lambda$ & $\alpha$ & $\Lambda_{\rm QCD}$ & $\gamma_u(=\gamma_d)$ & $\gamma_s$ \\
\hline
$1.62$ & $0.305$ & $0.311$ & $0.50$ & $0.21$ & $0.06$ & $0.27$ & $0.527$ & $0.35$ \\
GeV & GeV & GeV & GeV & GeV$^2$ & --- & GeV & --- & --- \\
\hline
\end{tabular}}
\end{center}
\end{table}

Solving the Salpeter equation simultaneously provides the numerical wave functions and the 
corresponding meson mass spectrum. While the resulting spectrum (e.g., the charmonium spectrum in 
Ref.~\cite{Chang:2010}) is already in good agreement with the data from the Particle Data Group (PDG) \cite{PDG}, we fine-tune the free parameter $V_0$ of 
the potential so that the calculated masses reproduce the PDG values exactly. This calibrated 
mass set is then used throughout the numerical evaluation. The masses employed are collected in 
Table~\ref{tab:masses}.

\begin{table}[ht]
\begin{center}
\caption{Masses of the relevant mesons (in MeV).}\label{tab:masses}
\setlength{\tabcolsep}{3mm}{
\begin{tabular}{c c c}
\hline
Meson & $J^{PC}$ & Mass (MeV) \\
\hline
$D^0$ / $\bar{D}^0$      & $0^{-}$ & 1864.84 \\
$D^\pm$                   & $0^{-}$ & 1869.66 \\
$D_s^\pm$                 & $0^{-}$ & 1968.35 \\
$D^{*}(2007)^0$ / $\bar{D}^{*}(2007)^0$ & $1^{-}$ & 2006.85 \\
$D^{*}(2010)^\pm$     & $1^{-}$ & 2010.26 \\
$D_s^{*\pm}$          & $1^{-}$ & 2112.2 \\
\hline
$\psi(3770)$ & $1^{--}$ & 3773.7 \\
$\psi(4040)$ & $1^{--}$ & 4039.6 \\
$\psi(4160)$ & $1^{--}$ & 4191 \\
\hline
\end{tabular}}
\end{center}
\end{table}

\subsection{Decays of $\psi(3770)$}
\label{sec:psi3770}

The $\psi(3770)$ is a $1D$-dominated charmonium state. It has been the subject of extensive 
experimental investigation, notably by the BES~\cite{BES:2005psi,BES:2006psi,BES:2006hadron,BES:2006xsect,BESIII:2015DD} and CLEO~\cite{CLEO:2005pipi,CLEO:2005nonDD,CLEOc:2007Dbr,CLEOc:2010psi} 
collaborations, and the PDG assigns it a total width of $\Gamma_{\rm tot}=27.2\pm1.0$~MeV \cite{PDG}. Because the state 
sits only slightly above the $D\bar{D}$ threshold, its OZI-allowed hadronic decays are 
restricted to the two channels $D^0\bar{D}^0$ and $D^+D^-$. From the PDG we infer that the 
sum of the two corresponding partial widths is $25.3^{+2.0}_{-2.4}$~MeV.

In Ref.~\cite{Ke:2026gnk} we studied the annihilation decay of vector charmonia into a lepton pair 
employing the four wave functions $\varphi_1$–$\varphi_4$, finding that $\varphi_2$ yields the 
best description of the data. Accordingly, we first employ $\varphi_2$ in the present work to 
compute the $\psi(3770) \to D\bar{D}$ width. Fitting the calculated total width to the measured 
value $\Gamma = 25.3~\text{MeV}$ fixes $\gamma_u = 0.527$; $\gamma_s = 0.35$ is taken as 
discussed earlier. These two $\gamma$ parameters are kept fixed in all subsequent calculations. 
Table~\ref{tab:psi3770}, for example, compares the $\psi(3770) \to D^0\bar{D}^0$ and 
$\psi(3770) \to D^+D^-$ partial widths obtained with the four wave functions individually, with 
their sum $\Gamma_{\text{sum}}$ quoted in the last row.

\begin{table}[ht]
\begin{center}
\caption{Strong decay widths of the $\psi(3770)$ computed with the four wave-function representations $\varphi_1$--$\varphi_4$ (in MeV).}\label{tab:psi3770}
\setlength{\tabcolsep}{2.5mm}{
\begin{tabular}{l c c c c c}
\hline
Channel & EX & $\varphi_1$ & $\varphi_2$ & $\varphi_3$ & $\varphi_4$ \\
\hline
$\Gamma_{D^0\bar{D}^0}$ & $14.1^{+1.6}_{-1.8}$ & $14.6^{+2.9}_{-2.7}$ & $14.4^{+5.0}_{-4.4}$ & $17.3^{+3.6}_{-3.2}$ & $10.2^{+1.8}_{-1.7}$\\
$\Gamma_{D^+D^-}$ & $11.2\pm1.5$ & $11.0^{+2.2}_{-2.1}$ & $10.9^{+3.9}_{-3.4}$ & $13.0^{+2.7}_{-2.5}$ & $7.84^{+1.5}_{-1.4}$\\
\hline
$\Gamma_{\text{sum}}$ & $25.3^{+2.0}_{-2.4}$ & $25.6^{+5.1}_{-4.7}$ & $25.3^{+8.9}_{-7.8}$ & $30.3^{+6.3}_{-5.7}$ & $18.1^{+3.2}_{-3.1}$\\
\hline
\end{tabular}}
\end{center}
\end{table}

The theoretical uncertainties are also listed in Table~\ref{tab:psi3770}. They are obtained by randomly 
varying all input parameters entering the calculation, the potential parameters and the quark masses 
among them, within $\pm 5\%$ of their central values; the resulting spread of the 
prediction is assigned as the uncertainty. One sees from the table that $\varphi_1$ yields a 
strong decay width of the $\psi(3770)$ equal to $25.6^{+5.1}_{-4.7}~\text{MeV}$, in remarkable 
agreement with both the $\varphi_2$ prediction and the data. We therefore conclude that the 
relativistic pieces contained in $\varphi_1$ but not in $\varphi_2$, specifically the $f_7$ and 
$f_8$ terms, contribute negligibly to the strong decay of the $\psi(3770)$. These very same 
terms, however, are known to play an important role in the annihilation decay of the 
$\psi(3770)$ into a lepton pair ~\cite{Ke:2026gnk}.
This difference can be understood as follows. The strong 
decay is overwhelmingly governed by the $D$-wave part of the wave function, so that the $P$-wave-like 
relativistic corrections ($f_7$ and $f_8$ terms) are suppressed relative to the leading $D$-wave contribution. In the 
electromagnetic annihilation channel, by contrast, the leading contribution is of $S$-wave nature; 
since the $S$-wave fraction in the $\psi(3770)$ is small, the $P$-wave terms become comparatively prominent.

It is worth noting that the uncertainty of $\varphi_2$ is clearly larger than that of $\varphi_1$: the total width is $25.3^{+8.9}_{-7.8}$~MeV for the former and $25.6^{+5.1}_{-4.7}$~MeV for the latter (the relative uncertainties are $33\%$ and $19\%$, respectively). The two agree within errors, but $\varphi_2$ is more sensitive to the parameters.
The upper two panels of Fig.~\ref{fig:wf} display the four independent radial wave functions of 
 the $\psi(3770)$ wave functions $\varphi_1$ and $\varphi_2$. The components $f_3$ and 
$f_4$ are of essentially $1D$-wave character and are nodeless. In $\varphi_1$, the $S$-wave 
pieces $f_5$ and $f_6$ are likewise nodeless and correspond to a $1S$ configuration; in 
$\varphi_2$, by contrast, $f_5$ and $f_6$ carry a single node and are therefore of $2S$ type. 
In the process $\psi(3770)\to D\bar D$, Eq.~\eqref{genamp} shows that the
transition amplitude is the overlap integral of the wave functions of the
three mesons involved in the initial and final states. Both final-state
mesons, $D$ and $\bar D$, are ground-state particles, and neither of their
wave functions contains a node. The $2S$ component of the $\psi(3770)$ wave
function, in contrast, contains one node. In the overlap integral, the portions of this $2S$ wave function on opposite sides of 
the node contribute to the transition amplitude with opposite signs. A parameter-induced shift of 
the node therefore modifies the magnitude of the amplitude rather sensitively. It is this sensitivity that 
makes the uncertainty on the decay width evaluated with $\varphi_2$ appreciably larger than the 
one obtained with $\varphi_1$. In other words,
although the $2S$ component carries only a small weight in $\psi(3770)$, it
governs the cancellation between contributions of opposite sign in the
overlap integral and is therefore responsible for the large uncertainty. The same mechanism operates for the $\psi(4040)$ and $\psi(4160)$: 
all partial-wave components exhibit nodal structures,  some even with more than one 
node, which accounts for the comparatively large uncertainties in these channels.

Compared with $\varphi_2$, the wave function $\varphi_3$ lacks the $f_1$ and $f_2$ terms but 
contains the $f_7$ and $f_8$ terms instead. All four of these are relativistic $P$-wave terms. 
However, the results show that the contributions of $f_1$ and $f_2$ to the decay width are 
significantly larger than those of $f_7$ and $f_8$. The $\psi(3770) \to D\bar{D}$ decay width 
evaluated with $\varphi_3$ is $30.3^{+6.3}_{-5.7}$ MeV, whose central value is $5.0~\text{MeV}$ larger 
than the $\varphi_2$ result, a noticeable difference. This indicates that the relativistic 
correction terms $f_1$ and $f_2$ cannot be neglected.

In contrast to the first three wave functions, $\varphi_4$ retains the relativistic correction 
sector ($f_1$–$f_4$) but lacks the independent $S$-wave pieces $f_5$ and $f_6$. Consequently, the 
solutions of the Salpeter equation built from $\varphi_4$ yield eigenvalues and wave functions 
that describe exclusively $D$-wave-dominated states, $1D$, $2D$, $3D$, …, and contain no $S$-wave-dominated 
states such as $1S$, $2S$, $3S$, … . The corresponding decay width comes out to be 
$18.1^{+3.2}_{-3.1}~\text{MeV}$, markedly below the predictions of the first three wave functions.

\subsection{Decays of $\psi(4040)$}
\label{sec:psi4040-results}

The $\psi(4040)$ is a vector charmonium whose dominant component is the $3S$ wave. 
The lower two panels of Fig.~\ref{fig:wf} display the wave functions for this state obtained with 
the $\varphi_1$ and $\varphi_2$ ansätze. In both cases the $S$-wave pieces $f_5$ and $f_6$ exhibit 
two nodes, the hallmark of a $3S$-dominated state. With its mass lying above the thresholds of $D\bar{D}$, $D^*\bar{D}$, $D^*\bar{D}^*$, and $D_s\bar D_s$, the $\psi(4040)$ can decay strongly into all of these 
channels via OZI-allowed transitions. The partial widths predicted by $\varphi_1$, $\varphi_2$, 
and $\varphi_3$ (the $\varphi_4$ does not produce a $3S$ solution and is therefore omitted) are listed in Table~\ref{tab:psi4040}, with ``c.c.'' indicating the charge-conjugate 
mode. The sums of the partial widths are quoted at the bottom of the table.

\begin{table}[ht]
\begin{center}
\caption{Strong decay widths of $\psi(4040)$ (in MeV) computed with $\varphi_1$--$\varphi_3$. The $\varphi_4$ does not produce a $3S$ dominant solution.}\label{tab:psi4040}
\setlength{\tabcolsep}{2.5mm}{
\begin{tabular}{l c c c}
\hline
Channel & $\varphi_1$ & $\varphi_2$ & $\varphi_3$ \\
\hline
$\Gamma_{D^0\bar{D}^0}$ & $0.045^{+0.65}_{-0.045}$ & $0.15^{+0.88}_{-0.15}$ & $(2.2^{+260}_{-2.2})\times10^{-3}$\\
$\Gamma_{D^+D^-}$ & $(1.7^{+3500}_{-1.7})\times10^{-4}$ & $(3.3^{+490}_{-3.3})\times10^{-3}$ & $0.012^{+0.30}_{-0.012}$\\
$\Gamma_{D^*(2007)^0\bar D^0+\text{c.c.}}$ & $32.7^{+1.2}_{-2.2}$ & $8.58^{+11}_{-3.8}$ & $63.4^{+2.1}_{-5.5}$\\
$\Gamma_{D^*(2010)^+D^-+\text{c.c.}}$ & $36.9^{+1.6}_{-2.3}$ & $10.1^{+13}_{-4.4}$ & $70.4^{+1.7}_{-4.4}$\\
$\Gamma_{D^*(2007)^0\bar D^*(2007)^0}$ & $34.0^{+11}_{-9.2}$ & $34.2^{+11}_{-9.2}$ & $28.8^{+13}_{-8.9}$\\
$\Gamma_{D^*(2010)^+D^*(2010)^-}$ & $23.8^{+8.0}_{-6.6}$ & $24.2^{+8.1}_{-6.6}$ & $20.3^{+9.5}_{-6.4}$\\
$\Gamma_{D_s^+D_s^-}$ & $8.50^{+0.78}_{-0.83}$ & $16.6^{+1.7}_{-4.0}$ & $5.36^{+2.8}_{-0.86}$\\
\hline
$\Gamma_{\text{sum}}$ & $136^{+19}_{-17}$ & $94^{+34}_{-23}$ & $188^{+16}_{-17}$\\
\hline
\end{tabular}}
\end{center}
\end{table}

For $\psi(4040)$, the summed widths given by the three representations differ considerably. The $\varphi_2$ representation gives $\Gamma_{\text{sum}} = 94^{+34}_{-23}$~MeV, in agreement with the experimental value $84.5 \pm 12.3$~MeV within errors \cite{PDG}. $\varphi_1$ overestimates the width ($136^{+19}_{-17}$~MeV) and $\varphi_3$ gives an even larger value ($188^{+16}_{-17}$~MeV). 
All three ansätze, $\varphi_1$, $\varphi_2$, and $\varphi_3$, correctly capture the strong 
suppression of the $D\bar{D}$ channel \cite{CLEO:2008ojp}. With $\varphi_2$, the hierarchy is particularly clear: 
the $D^*\bar{D}^*$ mode dominates at $58^{+19}_{-16}$~MeV, followed by 
$D^*\bar{D} + \text{c.c.}$ ($18.7^{+24}_{-8.2}$~MeV) and the strange channel 
$D_s^+\bar{D}_s^-$ ($16.6^{+1.7}_{-4.0}$~MeV), whereas the $D\bar{D}$ width is heavily suppressed, 
$0.15^{+1.4}_{-0.15}$~MeV.

For the decay channel $\psi(4040)\to D\bar{D}$, the phase space is sufficiently large, yet the partial width is very small, which is attributed to the two-node structure of the $\psi(4040)$ wave function. The wave function has opposite signs on either side of a node, so the contributions of the regions before and after the node to the amplitude subtract from each other. Our results show that this process takes place where the total amplitude is close to zero as a result of the node structure. This situation implies that the width must be extremely sensitive to the choice of parameters, since a change in the parameters shifts the node position and thereby alters the degree of cancellation in the amplitude. Moreover, within the range over which the node moves, the sign of the wave function changes and yields contributions of the opposite sign, causing the width to vary appreciably; this is the reason why the $\psi(4040)\to D\bar{D}$ process in Table~\ref{tab:psi4040} carries such a large error.

\begin{table}[ht]
\begin{center}
\caption{Decay width ratios of $\psi(4040)$.}\label{tab:psi4040_ratios}
\setlength{\tabcolsep}{2mm}{
\begin{tabular}{c c c c c}
\hline
Ratio & EX & $\varphi_1$ & $\varphi_2$ & $\varphi_3$ \\
\hline
$\frac{\Gamma_{D \bar D}}{\Gamma_{D^* \bar D +\text{c.c.}}}$ & $0.24\pm0.05\pm0.12$ \cite{BaBar:2009elc} & $(6.5^{+150}_{-3.1})\times10^{-4}$ & $(8.2^{+44}_{-6.2})\times10^{-3}$ & $(1.07^{+36}_{-0.30})\times10^{-4}$\\
$\frac{\Gamma_{D^0\bar{D}^0}}{\Gamma_{D^{*}(2007)^{0}\bar{D}^0+\text{c.c.}}}$ & $0.05\pm0.03$ \cite{Goldhaber:1977qn} & $(1.4^{+21}_{-1.4})\times10^{-3}$ & $0.018^{+0.065}_{-0.018}$ & $(3.5^{+450}_{-3.5})\times10^{-5}$\\
$\frac{\Gamma_{D^{*}(2010)^{+}D^-+\text{c.c.}}}{\Gamma_{D^{*}(2007)^{0}\bar{D}^0+\text{c.c.}}}$ & $0.95\pm0.09\pm0.10$ \cite{BaBar:2009elc} & $1.130^{+0.036}_{-0.029}$ & $1.180^{+0.024}_{-0.025}$ & $1.110^{+0.029}_{-0.023}$\\
$\frac{\Gamma_{D^* \bar D^*}}{\Gamma_{D^* \bar D+\text{c.c.}}}$ & $0.18\pm0.14\pm0.03$ \cite{BaBar:2009elc} & $0.83^{+0.31}_{-0.22}$ & $3.1^{+1.6}_{-1.5}$ & $0.37^{+0.21}_{-0.11}$\\
$\frac{\Gamma_{D^{*}(2007)^{0}\bar{D}^{*}(2007)^{0}}}{\Gamma_{D^{*}(2007)^{0}\bar{D}^{0}+\text{c.c.}}}$ & $32.0\pm12.0$ \cite{Goldhaber:1977qn} & $1.04^{+0.41}_{-0.28}$ & $4.0^{+2.1}_{-2.0}$ & $0.46^{+0.27}_{-0.14}$\\
\hline
\end{tabular}}
\end{center}
\end{table}

Table~\ref{tab:psi4040_ratios} lists several ratios of partial widths
together with the available experimental values. In the first two rows all
three representations predict $\Gamma_{D\bar D}/\Gamma_{D^*\bar D}\ll 1$,
i.e.\ a strong suppression of the $D\bar D$ channel, and the experimental results lead to the same
conclusion. The numerical values, however, differ to some extent: only the
ratio obtained with $\varphi_2$,
$\Gamma_{D^0\bar D^0}/\Gamma_{D^{*}(2007)^0\bar D^0+\text{c.c.}}
=0.018^{+0.065}_{-0.018}$, is consistent, within uncertainties, with the
measured value $0.05\pm0.03$. 
The relative uncertainties of these two measurements, about $54\%$ and $60\%$
(statistical and systematic errors added in quadrature), limit their
discriminating power.
Those of the last two are $79\%$ and
$38\%$ and are also large; their central values even point in the
opposite direction, so that their reference value is limited. 
The most constraining measurement is the third ratio,
$\Gamma_{D^{*+}D^-+\text{c.c.}}/\Gamma_{D^{*0}\bar D^0+\text{c.c.}}=0.95\pm0.09\pm0.10$, with a
relative uncertainty of only about $14\%$; it is also the ratio for which
all three representations have the smallest relative uncertainties
($2\%$--$3\%$), and all three predictions are consistent with the measurement within
$2\sigma$.

All uncertainties quoted in Table~\ref{tab:psi4040} come from the same parameter
scan. For each point in the parameter space we first compute the partial
widths of all channels and sum them to obtain the total width, and only then
take the maximum and minimum of the total width over the scan as its upper
and lower bounds; we do not first take the extrema of the individual
channels and add them afterwards. The ratios in Table~\ref{tab:psi4040_ratios} are
treated in the same way: the numerator and the denominator are always
evaluated at the same parameter point, and the extrema are taken of the
ratio itself rather than of the two separately. The correlations between the
parameters are thus retained along the entire calculation chain, so that the
uncertainty of the total width is smaller than the sum of the individual
uncertainties, and the bounds of the ratios are not overestimated by taking
the extrema of the numerator and the denominator independently. The same
procedure is used for all the uncertainties quoted in the remainder of this
paper and will not be repeated.

\subsection{Decays of $\psi(4160)$}
\label{sec:psi4160}

$\psi(4160)$ is slightly heavier than $\psi(4040)$, with a mass of
$4191$~MeV, which opens additional OZI-allowed strong decay channels, in
particular those containing $D_s^{*}D_s$. Unlike $\psi(4040)$, which is
predominantly a $3S$ state, $\psi(4160)$ is predominantly a $2D$ state and
can therefore also be described by the wave-function representation
$\varphi_4$. In Table~\ref{tab:psi4160} we list the OZI-allowed strong decay
widths of $\psi(4160)$ calculated with the four representations
$\varphi_1$--$\varphi_4$. The last row of the table gives the sum of these
partial widths. Compared with the measured total width of $\psi(4160)$,
$69\pm10$~MeV  \cite{PDG}, the values obtained with $\varphi_1$ and $\varphi_2$,
$112^{+40}_{-30}$~MeV and $100^{+23}_{-20}$~MeV, have lower bounds of $82$
and $80$~MeV that lie close to the upper end of the measured range,
$79$~MeV, and may be regarded as reasonably good results, whereas the
$\varphi_3$ value, $187^{+39}_{-42}$~MeV, exceeds the experimental one.

\begin{table}[ht]
\begin{center}
\caption{Strong decay widths of $\psi(4160)$ (in MeV).}\label{tab:psi4160}
\setlength{\tabcolsep}{2.5mm}{
\begin{tabular}{l c c c c}
\hline
Channel & $\varphi_1$ & $\varphi_2$ & $\varphi_3$ & $\varphi_4$ \\
\hline
$\Gamma_{D^0\bar{D}^0}$ & $11.2^{+5.3}_{-4.9}$ & $3.6^{+5.0}_{-1.6}$ & $18.8^{+4.6}_{-4.0}$ & $36.3^{+6.0}_{-7.9}$\\
$\Gamma_{D^+D^-}$ & $10.1^{+5.3}_{-4.7}$ & $3.3^{+4.7}_{-1.5}$ & $17.6^{+4.7}_{-3.9}$ & $36.1^{+6.6}_{-8.3}$\\
$\Gamma_{D^*(2007)^0\bar D^0+\text{c.c.}}$ & $0.050^{+0.92}_{-0.050}$ & $3.7^{+2.6}_{-2.5}$ & $(7.3^{+270}_{-7.3})\times10^{-3}$ & ---\\
$\Gamma_{D^*(2010)^+D^-+\text{c.c.}}$ & $(7.6^{+5700000}_{-5.6})\times10^{-7}$ & $2.9^{+2.4}_{-2.1}$ & $0.015^{+0.22}_{-0.015}$ & ---\\
$\Gamma_{D^*(2007)^0\bar D^{*}(2007)^0}$ & $39^{+15}_{-11}$ & $34.0^{+7.8}_{-6.2}$ & $70^{+17}_{-20}$ & ---\\
$\Gamma_{D^*(2010)^+D^{*}(2010)^-}$ & $37^{+14}_{-11}$ & $32.7^{+6.8}_{-5.5}$ & $70^{+16}_{-20}$ & ---\\
$\Gamma_{D_s^+D_s^-}$ & $2.9^{+1.7}_{-1.4}$ & $0.86^{+0.57}_{-0.33}$ & $1.41^{+1.1}_{-0.91}$ & $5.1^{+3.9}_{-2.9}$\\
$\Gamma_{D_s^{*+}D_s^-+\text{c.c.}}$ & $11.50^{+0.58}_{-0.76}$ & $19.00^{+0.62}_{-4.1}$ & $8.81^{+2.5}_{-0.44}$ & ---\\
\hline
$\Gamma_{\text{sum}}$ & $112^{+40}_{-30}$ & $100^{+23}_{-20}$ & $187^{+39}_{-42}$ & ---\\
\hline
\end{tabular}}
\end{center}
\end{table}

No results are given in the $\varphi_4$ column for final states containing
vector mesons, because these final-state vector mesons are predominantly
$S$-wave states while, as noted above, $\varphi_4$ can describe only
$D$-wave-dominant states; unlike the other three representations, it cannot
treat both $S$-wave- and $D$-wave-dominant states. Only the channels whose
final states consist of $S$-wave-dominant states have therefore been
computed. The $D^{*}D^{*}$ channels, which give the largest contributions in
the other three representations, have not been calculated; even so, the
three channels listed already sum to $77^{+16}_{-19}$~MeV, slightly above
the measured central value, so that the complete sum would clearly be much
larger than experiment.

$\psi(4160)$ is predominantly a $2D$ state, and its wave function contains
one node. Consequently, as for $\psi(4040)$, the presence of this node
suppresses the widths of some of the $\psi(4160)$ decay channels. As can be
seen from Table~\ref{tab:psi4160}, for the representations $\varphi_1$ and
$\varphi_3$ the channel $\psi(4160)\to D^*\bar D$ is strongly suppressed,
whereas for $\varphi_2$ the smallest contribution comes from
$\psi(4160)\to D_s^+D_s^-$, and the $D^*\bar D$ and $D\bar D$ widths are
comparable in size and also rather small. The $D^*\bar D^*$ channel, by
contrast, has the largest branching fraction for all three representations
$\varphi_1$, $\varphi_2$ and $\varphi_3$.

\begin{table}[ht]
\begin{center}
\caption{Decay width ratios of $\psi(4160)$.}\label{tab:psi4160_ratios}
\setlength{\tabcolsep}{2.5mm}{
\begin{tabular}{c c c c c}
\hline
Ratio & EX \cite{BaBar:2009elc} & $\varphi_1$ & $\varphi_2$ & $\varphi_3$ \\
\hline
$\frac{\Gamma_{D \bar D}}{\Gamma_{D^* \bar D^*}}$ & $0.02\pm0.03\pm0.02$  & $0.281^{+0.061}_{-0.064}$ & $0.104^{+0.13}_{-0.035}$ & $0.259^{+0.10}_{-0.075}$\\
$\frac{\Gamma_{D^* \bar D +\text{c.c.}}}{\Gamma_{D^* \bar D^*}}$ & $0.34\pm0.14\pm0.05$ & $(6.5^{+140}_{-3.2})\times10^{-4}$ & $0.099^{+0.081}_{-0.068}$ & $(1.590^{+33}_{-0.063})\times10^{-4}$\\
\hline
\end{tabular}}
\end{center}
\end{table}

Table~\ref{tab:psi4160_ratios} compares the two ratios with the BaBar
data. Deviations are quoted as the difference between the central values
divided by the quadrature sum of the experimental uncertainty and the
theoretical uncertainty on the side of the deviation; the spread of the
predictions over the parameter space is treated as a $1\sigma$ uncertainty.
The first experimental ratio is consistent with zero ($0.6\sigma$) and is best read as an
upper bound, $\Gamma_{D\bar D}/\Gamma_{D^*\bar D^*}<0.09$ at the $95\%$
confidence level. All three representations lie above it: $\varphi_2$
($0.104$) is the closest, at $1.7\sigma$, while $\varphi_1$ ($0.281$) and
$\varphi_3$ ($0.259$) are disfavoured at the $3.6\sigma$ and $2.9\sigma$
level, respectively. The second ratio, $0.34\pm0.14\pm0.05$, is more
discriminating: $\varphi_2$ gives $0.099$ ($1.4\sigma$), whereas $\varphi_1$
and $\varphi_3$ lie $500$ and $2100$ times below the data (about
$2.3\sigma$ in both cases). Both ratios therefore point to $\varphi_2$, the
only representation lying within $2\sigma$ of both measurements. Given the
limited precision of the data, a null measurement and a $44\%$
uncertainty, this comparison can discriminate only between predictions
that differ by orders of magnitude, and cannot yet test the $10\%$ level.

\subsection{Systematic comparison of the wave function representations}
\label{sec:comparison}

With all parameters fixed, the four representations can be compared on the
same footing. Since $\gamma_u$ was adjusted to reproduce the measured
$\psi(3770)$ width with $\varphi_2$, that entry is a calibration rather than
a prediction; the $\psi(3770)$ results for the other representations, and
all results for $\psi(4040)$ and $\psi(4160)$, are genuine predictions.

\begin{enumerate}
\item $\varphi_2$ is the only representation that describes all three
states: it predicts $94^{+34}_{-23}$~MeV for $\psi(4040)$, within
$0.4\sigma$ of $84.5\pm12.3$~MeV, and $100^{+23}_{-20}$~MeV for $\psi(4160)$,
within $1.4\sigma$ of $69\pm10$~MeV. It is also the only one whose ratios
agree with the data, matching
$\Gamma_{D^0\bar D^0}/\Gamma_{D^{*0}\bar D^0}=0.018^{+0.065}_{-0.018}$
against $0.05\pm0.03$ and lying within $2\sigma$ of both $\psi(4160)$
ratios. This is consistent with the dilepton analysis \cite{Ke:2026gnk},
where $\varphi_2$ likewise performed best, so that the electromagnetic and
strong decay channels select the same representation.

\item $\varphi_1$ reproduces the $\psi(3770)$ width
($25.6^{+5.1}_{-4.7}$~MeV) and is as close as $\varphi_2$ for $\psi(4160)$
($1.4\sigma$), but it overshoots $\psi(4040)$ by $2.5\sigma$
($136^{+19}_{-17}$~MeV). Retaining all eight components therefore does not
improve the description.

\item $\varphi_3$, which keeps $f_7$ and $f_8$ but drops $f_1$ and $f_2$,
gives the largest widths: $188^{+16}_{-17}$~MeV for $\psi(4040)$
($5.1\sigma$) and $187^{+39}_{-42}$~MeV for $\psi(4160)$ ($2.7\sigma$), and
its $\psi(4160)$ ratio $0.259$ is disfavoured at $2.9\sigma$. Since
$\varphi_1$ contains both groups of terms and lies in between, the $f_1$ and
$f_2$ terms are the ones that matter for the strong decay.

\item $\varphi_4$ has no independent $S$-wave pieces and describes
$D$-wave-dominated states only. It gives $18.1^{+3.2}_{-3.1}$~MeV for
$\psi(3770)$, about $2\sigma$ below the data, which shows that the $S$-wave
component is indispensable even for a $1D$-dominated state. Its spectrum
contains no $S$-wave-rich eigenstate, and channels with vector mesons cannot
be computed, so that the $\varphi_4$ column of Table~\ref{tab:psi4160} is
incomplete.
\end{enumerate}

These results point to the following picture. To describe $\psi(3770)$,
$\psi(4040)$ and $\psi(4160)$ simultaneously, the relativistic wave function
must contain independent $S$-wave, $P$-wave and $D$-wave components, because
$\psi(3770)$ is a $1D$-dominated state, $\psi(4040)$ is $3S$-dominated and
$\psi(4160)$ is $2D$-dominated. In every case the $P$-wave components provide
the dominant relativistic correction. Among the four choices, the
representation $\varphi_2$, which contains independent $S$- and $D$-wave
components together with the appropriate $P$-wave ones, is the only one
that describes all three vector states successfully at the same time.
Moreover, the wave functions of all these states contain nodes, including
the $2S$ component of $\psi(3770)$, and it is this nodal structure that
gives rise to the relatively large uncertainties of the theoretical
predictions, which in turn implies that more precise measurements will help
to discriminate among the descriptions.

\subsection{Comparison with other theoretical calculations}
\label{sec:comparison2}

In this subsection we compare the results of the $\varphi_2$ representation with other theoretical calculations in the literature. Tables~\ref{tab:compare3770}--\ref{tab:compare4160} summarize the comparison for the three states.

\begin{table}[ht]
\begin{center}
\caption{Decay widths of $\psi(3770)$ (in MeV) compared with other theoretical calculations. }\label{tab:compare3770}
\setlength{\tabcolsep}{3mm}{
\begin{tabular}{c c}
\hline
Source & $D\bar D$ \\
\hline
This work ($\varphi_2$) & $25.3^{+8.9}_{-7.8}$ \\
Fu \& Jiang 2019 (coupled channel) \cite{Fu:2018yxq} & 18.4 \\
Man \textit{et al.} 2024 (coupled channel) \cite{Man:2024} & 22.5 \\
Segovia \textit{et al.} 2013 (${}^3P_0$) \cite{Segovia:2013b} & 26.47 \\
Swanson (2006, ${}^3P_0$) \cite{Swanson:2006st} & 33 \\
Swanson (2006, $\rho K\rho$) \cite{Swanson:2006st} & 11 \\
\hline
Experiment \cite{PDG} & $25.3^{+2.0}_{-2.4}$ \\
\hline
\end{tabular}}
\end{center}
\end{table}

For $\psi(3770)$ in Table~\ref{tab:compare3770}, our $\varphi_2$ result $25.3^{+8.9}_{-7.8}$~MeV is consistent with the coupled-channel result (22.5~MeV) of Man \textit{et al.} \cite{Man:2024} and the ${}^3P_0$ prediction (26.47~MeV) of Segovia \textit{et al.} \cite{Segovia:2013b}, while the coupled-channel calculation of Fu \& Jiang \cite{Fu:2018yxq} gives a somewhat lower value of 18.4~MeV. All results lie within the range of the early Swanson estimate (11--33~MeV) \cite{Swanson:2006st}.

\begin{table}[ht]
\begin{center}
\caption{Decay widths of $\psi(4040)$ (in MeV) compared with other theoretical calculations.}\label{tab:compare4040}
\setlength{\tabcolsep}{1.5mm}{\footnotesize
\begin{tabular}{l c c c c c}
\hline
Source & $D\bar D$ & $D^*\bar D$+c.c. & $D^*\bar D^*$ & $D_s\bar D_s$ & $\Gamma_{\rm total}$ \\
\hline
This work ($\varphi_2$) & $0.15^{+1.4}_{-0.15}$ & $18.7^{+24}_{-8.2}$ & $58^{+19}_{-16}$ & $16.6^{+1.7}_{-4.0}$ & $94^{+34}_{-23}$ \\
Fu \& Jiang 2019 (coupled channel) \cite{Fu:2018yxq} & 2.75 & 21.4 & 56.6 & 6.15 & 86.9 \\
Man \textit{et al.} 2024 (coupled channel) \cite{Man:2024} & 6.4 & 9.7 & 21.8 & 0.1 & 38.2 \\
Segovia \textit{et al.} 2013 (${}^3P_0$) \cite{Segovia:2013b} & 4.61 & 22.23 & 82.35 & 2.08 & 111.27 \\
Gui \textit{et al.} 2018 (modified LP) \cite{Gui:2018} & --- & --- & --- & --- & $\sim$60 \\
Swanson (2006, ${}^3P_0$) \cite{Swanson:2006st} & 0.1 & 25 & 25 & 6 & 57 \\
Swanson (2006, $\rho K\rho$) \cite{Swanson:2006st} & 2.3 & 17 & 5 & 1.6 & 26 \\
\hline
Experiment \cite{PDG} & --- & --- & --- & --- & $84.5\pm12.3$ \\
\hline
\end{tabular}}
\end{center}
\end{table}

For $\psi(4040)$ in Table~\ref{tab:compare4040}, the most striking feature is that all calculations give a suppression of the $D\bar D$ channel. Our result $0.15^{+1.4}_{-0.10}$~MeV is consistent within errors with the early Swanson estimate of 0.1~MeV using  ${}^3P_0$ model \cite{Swanson:2006st}. The coupled-channel calculation of Fu \& Jiang \cite{Fu:2018yxq} gives a somewhat larger $D\bar D$ width of 2.75~MeV;  within the same model, Man \textit{et al.} recently report an even larger width of 6.4 MeV \cite{Man:2024}, where the suppression is less pronounced. Our total width $94^{+34}_{-23}$~MeV agrees with the experimental value $84.5\pm12.3$~MeV within errors, and the coupled-channel result 86.9~MeV by Fu \& Jiang, while the more recent coupled-channel calculation of Man \textit{et al.} \cite{Man:2024} gives a total width of 38.2~MeV. The ${}^3P_0$ result 111.27~MeV by Segovia \textit{et al.} \cite{Segovia:2013b} is higher than the experimental value. The result ($\sim$60~MeV) of the modified LP model \cite{Gui:2018}, where LP means potential-model based on a linear potential plus a Coulomb potential, and the Swanson estimate (57~MeV) \cite{Swanson:2006st} are on the low side but still compatible with experiment.

\begin{table}[ht]
\begin{center}
\caption{Decay widths of $\psi(4160)$ (in MeV) compared with other theoretical calculations.}\label{tab:compare4160}
\setlength{\tabcolsep}{2.5mm}{
\begin{tabular}{l c c c c c c}
\hline
Source & $D\bar D$ & $D^*\bar D$+c.c. & $D^*\bar D^*$ & $D_s\bar D_s$ & $D_s^*D_s$+c.c. & $\Gamma_{\rm total}$ \\
\hline
This work ($\varphi_2$) & $6.9^{+9.7}_{-3.1}$ & $6.6^{+5.0}_{-4.5}$ & $67^{+15}_{-12}$ & $0.86^{+0.57}_{-0.33}$ & $19.00^{+0.62}_{-4.1}$ & $100^{+23}_{-20}$ \\
Fu \& Jiang 2019 (coupled channel) \cite{Fu:2018yxq} & 11.7 & 7.98 & 61.4 & 0.03 & 8.75 & 89.9 \\
Man \textit{et al.} 2024 (coupled channel) \cite{Man:2024} & 11.8 & 13.2 & 20.0 & 0.2 & 1.0 & 46.3 \\
Segovia \textit{et al.} 2013 (${}^3P_0$) \cite{Segovia:2013b} & 22.82 & 2.22 & 83.73 & 0.24 & 6.94 & 115.95 \\
Swanson (2006, ${}^3P_0$) \cite{Swanson:2006st} & 12 & 0.3 & 27 & --- & --- & 57 \\
\hline
Experiment \cite{PDG} & --- & --- & --- & --- & --- & $69\pm10$ \\
\hline
\end{tabular}}
\end{center}
\end{table}

For $\psi(4160)$ in Table~\ref{tab:compare4160}, the $D^*\bar D^*$ channel is the dominant decay mode in all calculations. Our $D^*\bar D^*$ result $67^{+15}_{-12}$~MeV is consistent with the coupled-channel result 61.4~MeV by Fu \& Jiang \cite{Fu:2018yxq}. Our $\varphi_2$ total width $100^{+23}_{-20}$~MeV is consistent with the 89.9~MeV of Fu \& Jiang, lower than the ${}^3P_0$ prediction 115.95~MeV by Segovia \textit{et al.} \cite{Segovia:2013b}, and higher than the 46.3~MeV of Man \textit{et al.} \cite{Man:2024}. Compared with the experimental value $69\pm10$~MeV, our central value is higher by $31$~MeV, and the theoretical lower bound ($80$~MeV) nearly touches the experimental upper bound ($79$~MeV), the two overlapping only at the edge of their errors.

\subsection{The mixing angles}
\label{sec:mixing}

For these vector mesons, $S$--$D$ mixing is a well-known phenomenon, and it
is particularly important for the annihilation of $\psi(3770)$ and
$\psi(4160)$ into lepton pairs. The calculations of Man \textit{et al.}
\cite{Man:2024} show that such mixing has no pronounced effect on the
OZI-allowed strong decays of vector mesons. We nevertheless discuss it here,
because a satisfactory theoretical model can be established only by
reproducing the experimental results for the strong decays and for the
electromagnetic annihilation at the same time. Our approach shows that
these vector states are not merely $S$--$D$ mixed: their wave functions all
contain $S$--$P$--$D$ mixing.

From our previous study of the electromagnetic annihilation decays
\cite{Ke:2026gnk}, we obtain the $S:P:D$ ratios in the $\varphi_2$ wave
functions of $\psi(3770)$, $\psi(4040)$ and $\psi(4160)$: $0.139:0.221:1$,
$1:0.252:0.483$ and $0.495:0.339:1$, respectively. To allow a comparison
with other calculations, we drop the $P$-wave component, retain only the $S$
and $D$ parts, and define the mixing angle in the conventional way, with
$\tan\theta$ given by the ratio of the subleading to the leading amplitude;
this gives $\theta_{\psi(3770)}=7.92^{\circ}$,
$\theta_{\psi(4040)}=25.8^{\circ}$ and
$\theta_{\psi(4160)}=26.3^{\circ}$.

The small mixing angle obtained for $\psi(3770)$, $7.92^{\circ}$, agrees with
the early values of $10^{\circ}$ in Refs. \cite{Ding:1991vu,Kuang:1989ub},
as well as with the $10.5^{\circ}$ obtained recently by Man \textit{et al.} in a
coupled-channel calculation \cite{Man:2024}. The large mixing angles of
$25.8^{\circ}$ and $26.3^{\circ}$ for $\psi(4040)$ and $\psi(4160)$ are also
consistent with $21.2^{\circ}$ \cite{Zhao:2023hxc}, $23.4^{\circ}$ \cite{Man:2024}, and
$28^{\circ}$ and $29^{\circ}$ \cite{Badalian}, all of which are obtained
within coupled-channel models. In those works the mixing is induced by
meson-loop effects, whereas our approach yields a complete $S$--$P$--$D$
structure determined by the dynamics of the Salpeter equation; although the
underlying mechanisms are different, the results agree.

\section{Summary and conclusions}\label{sec:conclusion}

We have studied the OZI-allowed strong decays of the vector charmonia
$\psi(3770)$, $\psi(4040)$ and $\psi(4160)$ within the Salpeter equation,
using the four relativistic wave-function representations
$\varphi_1$--$\varphi_4$ and the revised ${}^3P_0$ model. The quark-pair
creation strength is fixed once, by the measured $\psi(3770)\to D\bar D$
width, and is not readjusted afterwards, so that all results for
$\psi(4040)$ and $\psi(4160)$ are parameter-free predictions.

For $\psi(3770)$, the representations $\varphi_1$, $\varphi_2$ and
$\varphi_3$ all reproduce the measured width, whereas $\varphi_4$, which
carries no independent $S$-wave component, falls short of it by about
$2\sigma$. For $\psi(4040)$, only $\varphi_2$ reproduces the measured total
width, $94^{+34}_{-23}$~MeV against $84.5\pm12.3$~MeV, while $\varphi_1$
overshoots it by $2.5\sigma$ and $\varphi_3$ by $5\sigma$. For $\psi(4160)$,
$\varphi_1$ and $\varphi_2$ give $112^{+40}_{-30}$ and
$100^{+23}_{-20}$~MeV, both about $1.4\sigma$ above the measured
$69\pm10$~MeV, while $\varphi_3$ gives $187^{+39}_{-42}$~MeV. The ratio
observables point in the same direction: $\varphi_2$ is the only
representation consistent with both $\psi(4160)$ ratios and the only one
reproducing
$\Gamma_{D^0\bar D^0}/\Gamma_{D^{*0}\bar D^0}=0.018^{+0.065}_{-0.018}$
against $0.05\pm0.03$, whereas the predictions of $\varphi_1$ and
$\varphi_3$ for the second $\psi(4160)$ ratio lie two to three orders of
magnitude away from the data.

The comparison of the four representations leads to a clear physical
picture. A successful relativistic description of vector quarkonia must
retain independent $S$-, $P$- and $D$-wave components, because the three
states are $1D$-, $3S$- and $2D$-dominated, respectively, and the $P$-wave
terms provide the dominant relativistic correction in every case. The
representation $\varphi_2$ satisfies this requirement, whereas $\varphi_1$
does not improve the description despite containing all eight components,
$\varphi_3$ drops the $f_1$ and $f_2$ terms that turn out to matter, and
$\varphi_4$ lacks the independent $S$-wave sector altogether.

The nodal structure of the wave functions accounts for the size of the
theoretical uncertainties: the transition amplitude is a sum of two or three
contributions of alternating sign, so that a parameter-induced shift of a
node changes the width appreciably. This is why the $\psi(3770)$ width
computed with $\varphi_2$ carries a $33\%$ uncertainty while $\varphi_1$
gives $19\%$, and why the $D\bar D$ channel of $\psi(4040)$ is so poorly
determined.

The mixing angles extracted from $\varphi_2$, namely $7.92^{\circ}$ for
$\psi(3770)$, $25.8^{\circ}$ for $\psi(4040)$ and $26.3^{\circ}$ for
$\psi(4160)$, agree with the values obtained in coupled-channel models,
although in the present approach they follow from the complete
$S$--$P$--$D$ structure dictated by the Salpeter equation rather than from
meson-loop mixing. More precise measurements, particularly of the
$\psi(4160)$ total width and of the decay ratios, will make it possible to
resolve the remaining degeneracy between $\varphi_1$ and $\varphi_2$.

{\bf Acknowledgments}

This work was supported by the National Natural Science Foundation of China (NSFC) under the Grants No. 12575097 and No. 12375085. T. Wang was also supported by the Fundamental Research Funds for the Central Universities (2023FRFK06009). Q. Li was supported by the National Key R\&D Program of China\,(2022YFA1604803) and the Natural Science Basic Research Program of Shaanxi\,(No.\,2025JC-YBMS-020). 

\appendix
{\allowdisplaybreaks[1]
\section{The Salpeter equation}
\label{app:salpeter}

The Bethe-Salpeter (BS) equation \cite{Salpeter:1951} is the relativistic
dynamical equation describing two-body bound states. In the instantaneous
approximation, that is, when the interaction kernel is assumed to be
independent of the time component, the BS equation reduces to the Salpeter
equation \cite{Salpeter:1952},
\begin{align}\label{Lorentz-covariant}
(M - w_1 - w_2) \varphi^{++}(q_{\perp}) &=
\Lambda_1^{+}(q_{\perp}) \left[ \int \frac{d\vec{k}}{(2 \pi)^{3}}
V({q}_{\perp}-k_{\perp}) \varphi({k}_{\perp})\right] \Lambda_2^{+}(q_{\perp}), \nonumber\\
(M + w_1 + w_2) \varphi^{--}(q_{\perp}) &=
-\Lambda_1^{-}(q_{\perp})\left[ \int \frac{d\vec{k}}{(2 \pi)^{3}}
V({q}_{\perp}-k_{\perp}) \varphi({k}_{\perp})\right] \Lambda_2^{-}(q_{\perp}), \\
\varphi^{+-}(q_{\perp}) &= \varphi^{-+}(q_{\perp}) = 0, \nonumber
\end{align}
where the positive- and negative-energy components are defined by the
projection operators
\begin{equation}\label{posi}
\varphi^{\pm\pm}(q_\perp) \equiv \Lambda_1^\pm(q_\perp) \frac{\slashed{P}}{M}
\varphi(q_\perp) \frac{\slashed{P}}{M} \Lambda_2^\pm(q_\perp),
\end{equation}
and $\Lambda_i^{\pm}$ are the positive- and negative-energy projection
operators,
\begin{equation}
\Lambda_{i}^{\pm}(q_{\perp}) =
\frac{1}{2\omega_i}\left[\frac{\slashed P}{M}\omega_i \pm
J(i)\left(m_i+\slashed q_{\perp}\right)\right],
\end{equation}
where $J(1) = +1$ corresponds to the quark and $J(2) = -1$ to the antiquark,
$\omega_i = \sqrt{m_i^2-q_{\perp}^2}=\sqrt{m_i^2+\vec{q}^{\,2}}$ is the quark
energy, $m_i$ is the constituent quark mass. The
normalization condition of the wave function is
\begin{equation}\label{normalization}
\int \frac{d\vec{q}}{(2 \pi)^{3}}
\operatorname{Tr}\left[\bar{\varphi}^{++}(q_{\perp}) \frac{\slashed P}{M}
\varphi^{++}(q_{\perp}) \frac{\slashed P}{M}
- \bar{\varphi}^{--} (q_{\perp}) \frac{\slashed P}{M} \varphi^{--}(q_{\perp})
\frac{\slashed P}{M}\right] = 2 M.
\end{equation}

In our calculation, the wave function is relativistic, to avoid double counting, a non-relativistic kernel is needed \cite{Wang:2019tqf}, we adopt the modified Cornell potential as the interaction kernel \cite{Wang:2005},
\begin{equation}\label{potential}
V(\vec{q}) = -\!\left(\frac{\lambda}{\alpha}+V_{0}\right)\!\delta^{3}(\vec{q})
+\frac{\lambda}{\pi^{2}}\frac{1}{(\vec{q}^{2}+\alpha^{2})^{2}}
-\gamma_{0}\otimes\gamma^{0}\frac{2}{3\pi^{2}}
\frac{\alpha_{s}(\vec{q})}{(\vec{q}^{2}+\alpha^{2})},
\end{equation}
where $\lambda$ is the string tension, $V_0$ is a free parameter used to
adjust the overall mass scale, and $\alpha$ is the screening parameter
regularizing the infrared divergence. The running strong coupling constant
$\alpha_s(\vec{q})$ in the Coulomb term is taken in the one-loop form
\cite{Richardson,Wang:2005},
\begin{equation}\label{runningalpha}
\alpha_s(\vec{q}) = \frac{4\pi}{9}
\frac{1}{\ln\!\big(e + \vec{q}^{\,2}/\Lambda_{\rm QCD}^{2}\big)},
\end{equation}
where $e=2.71828$ is the Euler number and $\Lambda_{\rm QCD}$ is the QCD
scale parameter. The factor $e$ inside the logarithm ensures that the
running coupling avoids the Landau pole in the infrared region.

\section{Constraint relations, normalization conditions, and
positive-energy coefficients}
\label{app:coeff}

\subsection{Constraint relations and normalization conditions}
\label{app:constraints}

This appendix collects the constraint relations and normalization
conditions for all four representations. All the equations
hold only for solutions satisfying the constraint relations $\varphi^{+-}=\varphi^{-+}=0$.

\subsubsection{The $\varphi_1$ representation}

The wave function $\varphi_1$ involves eight radial wave functions,
four of which are independent. We choose $f_3, f_4, f_5,$ and $f_6$
as the independent set; the remaining radial wave functions are fixed
by the constraint relations, whose explicit forms are given in the
main text. Here we only present the normalization condition:
\begin{equation}\label{phi1-norm}
\int \frac{d\vec{q}}{(2\pi)^3}\,
\frac{8 w_1 w_2 \left(M^2 q_{\perp}^2 (f_4 f_5+f_3 f_6)
+f_3 f_4 q_{\perp}^4+3 f_5 f_6 M^4\right)}
{3 M^3 (m_1 w_2+m_2 w_1)} = 1.
\end{equation}

\subsubsection{The $\varphi_2$ representation}

Choosing $f_3$, $f_4$, $f_5$, and $f_6$ as the independent functions:
\begin{align}\label{phi2constr-app}
f_1 = \frac{(w_1+w_2)(f_3 q_{\perp}^2+f_5 M^2)}{M(m_1 w_2+m_2 w_1)},&&
f_2 = \frac{(w_1-w_2)(f_4 q_{\perp}^2+f_6 M^2)}{M(m_1 w_2+m_2 w_1)}, && f_7 = 0, && f_8 = 0.
\end{align}
The normalization condition reads
\begin{equation}\label{phi2-norm-app}
\int \frac{d\vec{q}}{(2\pi)^3} \left[
\frac{8 w_1 w_2 (f_3 q_{\perp}^2+f_5 M^2)(f_6 M^2+f_4 q_{\perp}^2)}
{3 M^3 (m_1 w_2+m_2 w_1)}
+ \frac{4}{3} f_5 f_6 M \left(\frac{m_1}{w_1}+\frac{m_2}{w_2}\right)
\right] = 1.
\end{equation}

\subsubsection{The $\varphi_3$ representation}

Choosing $f_3$, $f_4$, $f_5$, and $f_6$ as the independent functions, $f_1$
and $f_2$ are set to zero, while $f_7$ and $f_8$ are determined by the
constraint conditions:
\begin{align}\label{phi3constr-app}
f_7 = -\frac{(f_3 q_{\perp}^2+3 f_5 M^2)(w_2-w_1)}{2 M (m_1 w_2+m_2 w_1)}, &&
f_8 = -\frac{f_6 M^2+f_4 q_{\perp}^2}{M(m_1+m_2)}
+\frac{(w_1+w_2)(3 f_6 M^2+f_4 q_{\perp}^2)}{2 M (m_1 w_2+m_2 w_1)}.
\end{align}
The normalization condition reads
\begin{multline}\label{phi3-norm}
\int \frac{d\vec{q}}{(2\pi)^3}\,
\frac{1}{3 M^3 q_{\perp}^2 w_1 w_2 (m_1+m_2)(w_1+w_2)}
\Bigl\{ f_3 f_4 q_{\perp}^4\Bigl[ \bigl(q_{\perp}^2-w_1 w_2+m_1 m_2\bigr)\mathcal{K}
+2 m_1 m_2 q_{\perp}^2(w_1+w_2)^2 \Bigr] \\
+ (f_4 f_5+f_3 f_6) M^2 q_{\perp}^2\Bigl[ 3 m_1 m_2\mathcal{K}
+q_{\perp}^2(w_1+w_2)^2\bigl(q_{\perp}^2-5 w_1 w_2\bigr)
+12 w_1^2 w_2^2\bigl(q_{\perp}^2-w_1 w_2\bigr) \Bigr] \\
+ f_5 f_6 M^4\Bigl[ 3 m_1 m_2\bigl(\mathcal{K}+8 w_1^2 w_2^2\bigr)
+q_{\perp}^2(w_1+w_2)^2\bigl(q_{\perp}^2-9 w_1 w_2\bigr)
+36 w_1^2 w_2^2\bigl(q_{\perp}^2-w_1 w_2\bigr) \Bigr] \Bigr\} = 1,
\end{multline}
where $\mathcal{K}\equiv 4w_1^2w_2^2+q_{\perp}^2(w_1+w_2)^2$.

\subsubsection{The $\varphi_4$ representation}

This representation has no independent $S$-wave term. Among the four
$P$-wave functions only $f_1$ and $f_2$ are retained, $f_7$ and $f_8$ are
set to zero, and $f_1$ and $f_2$ are again determined by the constraint
conditions:
\begin{align}\label{phi4constr-app}
f_1 &= \frac{f_3 q_{\perp}^2 (w_1+w_2)}{M(m_1 w_2+m_2 w_1)},\qquad
f_2 = \frac{f_4 q_{\perp}^2 (w_1-w_2)}{M(m_1 w_2+m_2 w_1)},\qquad
f_7 = 0,\qquad f_8 = 0.
\end{align}
The normalization condition simplifies to
\begin{equation}\label{phi4-norm}
\int \frac{d\vec{q}}{(2\pi)^3}\,
\frac{8 f_3 f_4 q_{\perp}^4 w_1 w_2}{3 M^3 (m_1 w_2+m_2 w_1)} = 1.
\end{equation}

\subsection{Positive-energy coefficients of the $1^-$ wave functions}
\label{app:coeff1}

The positive-energy coefficients $A_i$ of $\varphi_1$, $\varphi_2$, and
$\varphi_4$ are listed below; they follow directly from the definitions,
Eq.~(\ref{posi}) and Eq.~(\ref{positron}). 

\subsubsection{The $\varphi_1$ representation}

\begin{align}
A_1 &= \frac{q_{\perp}^2 [f_3 (w_1+w_2)+f_4 (m_1+m_2)]+2 f_5 M^2 w_2+2 f_6 M^2 m_2}{2 M (m_1 w_2+m_2 w_1)}, \nonumber\\
A_2 &= \frac{q_{\perp}^2 [f_3 (m_1-m_2)+f_4 (w_1-w_2)]-2 f_5 M^2 m_2-2 f_6 M^2 w_2}{2 M (m_1 w_2+m_2 w_1)}, \nonumber\\
A_3 &= \frac{1}{2}\left( f_3 + \frac{f_4 (m_1+m_2)}{w_1+w_2}
+ \frac{2 f_6 M^2}{m_1 w_2+m_2 w_1} \right), \qquad A_5 = \frac{1}{2}\left( f_5 + \frac{f_6 (w_1+w_2)}{m_1+m_2} \right),\nonumber\\
A_4 &= \frac{1}{2}\left( \frac{f_3 (w_1+w_2)}{m_1+m_2} + f_4
- \frac{2 f_5 M^2}{m_1 w_2+m_2 w_1} \right), \qquad A_6 = \frac{1}{2}\left( -f_6 - \frac{f_5 (m_1+m_2)}{w_1+w_2} \right), \nonumber\\
A_7 &= \frac{M [f_5 (w_1-w_2)-f_6 (m_2-m_1)]}{2 (m_1 w_2+m_2 w_1)}, \qquad
A_8 = \frac{M [-f_6 (w_1+w_2)-f_5 (m_1+m_2)]}{2 (m_1 w_2+m_2 w_1)}.
\end{align}

\subsubsection{The $\varphi_2$ representation}

\begin{align}
A_1 &= \frac{2 q_{\perp}^2 w_1 w_2 [f_3 (w_1+w_2)+f_4 (m_1+m_2)]
+f_5 M^2 [m_1 m_2 (w_2-w_1)+q_{\perp}^2 (w_1-w_2)+w_1 w_2 (w_1+3 w_2)]}
{4 M w_1 w_2 (m_1 w_2+m_2 w_1)} \nonumber\\
&\quad - \frac{f_6 M^2 [m_2 w_1 (w_1-3 w_2)-m_1 w_2 (w_1+w_2)]}
{4 M w_1 w_2 (m_1 w_2+m_2 w_1)}, \nonumber\\
A_2 &= \frac{2 q_{\perp}^2 w_1 w_2 [f_3 (m_1-m_2)+f_4 (w_1-w_2)]
-f_5 M^2 [m_1 w_2 (w_2-w_1)+m_2 w_1 (w_1+3 w_2)]}
{4 M w_1 w_2 (m_1 w_2+m_2 w_1)} \nonumber\\
&\quad - \frac{f_6 M^2 [m_1 m_2 (w_1+w_2)+q_{\perp}^2 (w_1+w_2)-w_1 w_2 (w_1-3 w_2)]}
{4 M w_1 w_2 (m_1 w_2+m_2 w_1)}, \nonumber\\
A_3 &= \frac{f_3}{2}+\frac{f_4 (m_1+m_2)}{2 (w_1+w_2)}
+\frac{M^2 (w_1-w_2)[f_5 (m_2-m_1)-f_6 (w_1-w_2)]}
{4 w_1 w_2 (m_1 w_2+m_2 w_1)}, \nonumber\\
A_4 &= \frac{f_3 (w_1+w_2)}{2 (m_1+m_2)}+\frac{f_4}{2}
+\frac{M^2 (w_1+w_2)[-f_6 (m_1+m_2)-f_5 (w_1+w_2)]}
{4 w_1 w_2 (m_1 w_2+m_2 w_1)}, \nonumber\\
A_5 &= \frac{m_1 w_2+m_2 w_1}{4 w_1 w_2}\, 
\frac{f_5 (w_1+w_2)+f_6 (m_1+m_2)}{m_1+m_2}, \qquad A_7= \frac{M [f_5 (m_1-m_2)-f_6 (w_2-w_1)]}{4 w_1 w_2},\nonumber\\
A_6 &= \frac{m_1 w_2+m_2 w_1}{4 w_1 w_2}\,
\frac{-f_6 (m_1+m_2)-f_5 (w_1+w_2)}{w_1+w_2}, \qquad 
A_8 = -\frac{M [f_6 (m_1+m_2)+f_5 (w_1+w_2)]}{4 w_1 w_2}.
\end{align}

\subsubsection{The $\varphi_3$ representation}

\begin{align}
A_1 ={}& \frac{1}{8M(m_1+m_2)q_{\perp}^{2}w_1w_2(w_1+w_2)}
\Bigl\{\, f_3(m_1+m_2)q_{\perp}^{2}
\bigl[(m_1+m_2)q_{\perp}^{2}(w_1+3w_2)-2m_1m_2(m_1w_2-m_2w_1)\bigr]
 \nonumber\\
&\quad + f_4q_{\perp}^{2}\Bigl[2m_1^2(m_1-m_2)m_2^2
+q_{\perp}^{2}\bigl((m_1+m_2)^3-2(m_1-m_2)m_2^2-4(m_1+m_2)q_{\perp}^{2}\bigr) \nonumber\\
&\quad -2m_1(m_1-m_2)m_2w_1w_2+4(m_1+m_2)q_{\perp}^{2}w_1w_2\Bigr]
- f_5M^2(m_1+m_2)\bigl[6m_1m_2(m_1w_2-m_2w_1) \nonumber\\
&\quad +q_{\perp}^{2}\bigl((3m_1-m_2)w_1-(3m_1+7m_2)w_2\bigr)\bigr]
+ f_6M^2\Bigl[\,6m_1^2(m_1-m_2)m_2^2 \nonumber\\
&\quad +q_{\perp}^{2}\bigl((7m_2-3m_1)(m_1^2+m_2^2)-8m_2q_{\perp}^{2}\bigr)
-6m_1(m_1-m_2)m_2w_1w_2+8m_2q_{\perp}^{2}w_1w_2\Bigr]\Bigr\},\nonumber
\\[1mm]
A_2 ={}& \frac{1}{8M(m_1+m_2)q_{\perp}^{2}w_1w_2(w_1+w_2)}
\Bigl\{\, f_3(m_1+m_2)q_{\perp}^{2}\bigl[2m_1^2m_2^2+(m_1^2-3m_2^2)q_{\perp}^{2}-2m_1m_2w_1w_2\bigr] \nonumber\\
&\quad + f_4(m_1-m_2)q_{\perp}^{2}
\bigl[(m_1+3m_2)q_{\perp}^{2}(w_1+w_2)-2m_1m_2(m_1w_2-m_2w_1)\bigr]
+ f_5M^2(m_1+m_2)\bigl[6m_1^2m_2^2 \nonumber\\
&\quad -3m_1^2q_{\perp}^{2}-7m_2^2q_{\perp}^{2}+4q_{\perp}^{4}
-(6m_1m_2+4q_{\perp}^{2})w_1w_2\bigr] - f_6M^2\Bigl[\,6m_1m_2(m_1-m_2)(m_1w_2-m_2w_1) \nonumber\\
&\quad +q_{\perp}^{2}\bigl((3m_1^2-6m_1m_2+11m_2^2-8q_{\perp}^{2})w_1 
 +(7m_1^2-6m_1m_2+7m_2^2-8q_{\perp}^{2})w_2\bigr)\Bigr]\Bigr\},\nonumber
\\[1mm]
A_3 ={}& \frac{1}{8(m_1+m_2)q_{\perp}^{2}w_1w_2}
\Bigl\{\, f_3(m_1+m_2)q_{\perp}^{2}\bigl(3m_1m_2+q_{\perp}^{2}+w_1w_2\bigr)
+ f_4q_{\perp}^{2}\bigl[4m_1m_2(w_1+w_2)\nonumber\\
&\quad +(m_1-m_2)(m_1w_2-m_2w_1)\bigr]
 + f_5M^2(m_1+m_2)\bigl(3m_1m_2+q_{\perp}^{2}-3w_1w_2\bigr)\nonumber\\
&\quad  + f_6M^2\bigl[4q_{\perp}^{2}(w_1+w_2)-3(m_1-m_2)(m_1w_2-m_2w_1)\bigr]\Bigr\},\nonumber
\\[1mm]
A_4 ={}& \frac{1}{8q_{\perp}^{2}w_1w_2(w_1+w_2)}
\Bigl\{\, f_3(m_1+m_2)q_{\perp}^{2}\bigl[3(m_1m_2-q_{\perp}^{2})+w_1w_2\bigr]
+ f_4q_{\perp}^{2}\bigl[4(m_1m_2-q_{\perp}^{2})(w_1+w_2) \nonumber\\
&\quad +(m_1-m_2)(m_1w_2-m_2w_1)\bigr]
+ 3f_5M^2(m_1+m_2)\bigl(m_1m_2-q_{\perp}^{2}-w_1w_2\bigr)\nonumber\\
&\quad- 3f_6M^2(m_1-m_2)(m_1w_2-m_2w_1)\Bigr\},\nonumber
\\[1mm]
A_5 ={}& \frac{1}{8M^2(m_1+m_2)w_1w_2}
\Bigl\{\, -f_3(m_1+m_2)q_{\perp}^{2}\bigl(m_1m_2-q_{\perp}^{2}-w_1w_2\bigr)
+ f_4(m_1-m_2)q_{\perp}^{2}(m_1w_2-m_2w_1) \nonumber\\
&\quad - f_5M^2(m_1+m_2)\bigl(m_1m_2-q_{\perp}^{2}-5w_1w_2\bigr) \nonumber\\
&\quad+ f_6M^2\bigl[4(m_1m_2-q_{\perp}^{2})(w_1+w_2)+5(m_1-m_2)(m_1w_2-m_2w_1)\bigr]\Bigr\},\nonumber
\\[1mm]
A_6 ={}& \frac{1}{8M^2w_1w_2(w_1+w_2)}
\Bigl\{\, f_3(m_1+m_2)q_{\perp}^{2}\bigl(m_1m_2-q_{\perp}^{2}-w_1w_2\bigr)
- f_4(m_1-m_2)q_{\perp}^{2}(m_1w_2-m_2w_1) \nonumber\\
&\quad + f_5M^2(m_1+m_2)\bigl(m_1m_2-q_{\perp}^{2}-5w_1w_2\bigr)\nonumber\\
&\quad- f_6M^2\bigl[4(m_1m_2-q_{\perp}^{2})(w_1+w_2)+5(m_1-m_2)(m_1w_2-m_2w_1)\bigr]\Bigr\},\nonumber
\\[1mm]
A_7 ={}& \frac{1}{8M(m_1+m_2)q_{\perp}^{2}w_1w_2(w_1+w_2)}
\Bigl\{\, f_3(m_1+m_2)q_{\perp}^{2}
\bigl[(m_1+m_2)q_{\perp}^{2}(w_1-w_2)+2m_1m_2(m_1w_2-m_2w_1)\bigr] \nonumber\\
&\quad - f_4(m_1-m_2)q_{\perp}^{2}
\bigl[2m_1^2m_2^2-(m_1^2+m_2^2)q_{\perp}^{2}-2m_1m_2w_1w_2\bigr]
+ f_5M^2(m_1+m_2)\Bigl[\,6m_1m_2(m_1w_2-m_2w_1) \nonumber\\
&\quad +q_{\perp}^{2}\bigl(3(m_1+m_2)(w_1-w_2)+2(m_1-m_2)(w_1+w_2)\bigr)\Bigr]
- f_6M^2(m_1-m_2)\bigl[6m_1^2m_2^2-5(m_1^2+m_2^2)q_{\perp}^{2} \nonumber\\
&\quad +4q_{\perp}^{4}-(6m_1m_2+4q_{\perp}^{2})w_1w_2\bigr]\Bigr\},\nonumber
\\[1mm]
A_8 ={}& \frac{1}{8M(m_1+m_2)q_{\perp}^{2}w_1w_2(w_1+w_2)}
\Bigl\{\, f_3(m_1+m_2)q_{\perp}^{2}
\bigl[2m_1^2m_2^2-(m_1^2+m_2^2)q_{\perp}^{2}-2m_1m_2w_1w_2\bigr] \nonumber\\
&\quad - f_4(m_1-m_2)q_{\perp}^{2}
\bigl[(m_1-m_2)q_{\perp}^{2}(w_1+w_2)+2m_1m_2(m_1w_2-m_2w_1)\bigr]
+ f_5M^2(m_1+m_2)\bigl[6m_1^2m_2^2 \nonumber\\
&\quad -5(m_1^2+m_2^2)q_{\perp}^{2}+4q_{\perp}^{4}
-(6m_1m_2+4q_{\perp}^{2})w_1w_2\bigr] - f_6M^2\Bigl[\,6m_1m_2(m_1-m_2)(m_1w_2-m_2w_1) \nonumber\\
&\quad +q_{\perp}^{2}\bigl((5m_1^2-6m_1m_2+9m_2^2-8q_{\perp}^{2})w_1
+(9m_1^2-6m_1m_2+5m_2^2-8q_{\perp}^{2})w_2\bigr)\Bigr]\Bigr\}.
\end{align}

\subsubsection{The $\varphi_4$ representation}

\begin{align}
A_1 &= \frac{q_{\perp}^2 [f_3 (w_1+w_2)+f_4 (m_1+m_2)]}{2 M (m_1 w_2+m_2 w_1)},  \qquad
A_2 = \frac{q_{\perp}^2 [f_3 (m_1-m_2)+f_4 (w_1-w_2)]}{2 M (m_1 w_2+m_2 w_1)}, \nonumber\\
A_3 &= \frac{f_3 m_1 w_2+f_3 m_2 w_1+f_4 m_1 m_2+f_4 q_{\perp}^2+f_4 w_1 w_2}
{2 (m_1 w_2+m_2 w_1)},\qquad A_5 = 0, \qquad A_6 = 0, \nonumber\\
A_4 &= \frac{f_3 m_1 m_2-f_3 q_{\perp}^2+f_3 w_1 w_2+f_4 m_1 w_2+f_4 m_2 w_1}
{2 (m_1 w_2+m_2 w_1)}, 
\qquad A_7 = 0, \qquad A_8 = 0. \nonumber
\end{align}

\subsection{Positive-energy coefficients of the $0^-$ wave function}
\label{app:coeff0}

For the positive-energy wave function of the pseudoscalar $0^-$ meson,
Eq.~(\ref{positive 0-}), the coefficients read
\begin{equation}\label{eq:Bcoeff}
\begin{aligned}
B_1 &= \frac{M}{2}\left(\frac{h_1(w_1+w_2)}{m_1+m_2}+h_2\right), &\quad
B_3 &= -\frac{B_1 M (w_1-w_2)}{m_1 w_2 + m_2 w_1}, \\
B_2 &= \frac{M}{2}\left(\frac{h_2(m_1+m_2)}{w_1+w_2}+h_1\right), &
B_4 &= -\frac{B_1 M (m_1+m_2)}{m_1 w_2 + m_2 w_1},
\end{aligned}
\end{equation}
where $h_1$ and $h_2$ denote the radial wave functions obtained from solving
the Salpeter equation.

}

\end{document}